\documentclass[bibyear]{aa}  
\usepackage{graphicx}
\usepackage{txfonts}
\usepackage{natbib,twoopt}
\AtBeginDocument{} 
\usepackage[breaklinks=true,colorlinks=true,allcolors=blue]{hyperref} 
\makeatletter
\newcommandtwoopt{\citeads}[3][][]{\href{http://adsabs.harvard.edu/abs/#3}%
{\def\hyper@linkstart##1##2{}%
\let\hyper@linkend\@empty\citealp[#1][#2]{#3}}}
\newcommandtwoopt{\citepads}[3][][]{\href{http://adsabs.harvard.edu/abs/#3}%
{\def\hyper@linkstart##1##2{}%
\let\hyper@linkend\@empty\citep[#1][#2]{#3}}}
\newcommandtwoopt{\citetads}[3][][]{\href{http://adsabs.harvard.edu/abs/#3}%
{\def\hyper@linkstart##1##2{}%
\let\hyper@linkend\@empty\citet[#1][#2]{#3}}}
\newcommandtwoopt{\citeyearads}[3][][]%
{\href{http://adsabs.harvard.edu/abs/#3}
{\def\hyper@linkstart##1##2{}%
\let\hyper@linkend\@empty\citeyear[#1][#2]{#3}}}
\makeatother
\bibpunct{(}{)}{;}{a}{}{,} 
\usepackage{xcolor}

\newcommand{\etal}{et al. }
\newcommand{\ha}{H$\alpha$}
\newcommand{\be}{\begin{equation}}
\newcommand{\ee}{\end{equation}}
\newcommand{\bea}{\begin{eqnarray}}
\newcommand{\eea}{\end{eqnarray}}

\begin{document} 

\authorrunning{C. E. Alissandrakis etal}
\title{Multi-wavelength Observations of a Metric Type-II Event}
\author{C. E. Alissandrakis\inst{1}, A. Nindos\inst{1}, S. Patsourakos\inst{1}
\and
A. Hillaris\inst{2}
}
\institute{Department of Physics, University of Ioannina, GR-45110 Ioannina, 
Greece\\
\email{calissan@uoi.gr, anindos@uoi.gr, spatsour@uoi.gr}
\and{Section of Astrophysics, Astronomy \& Mechanics, Department of Physics, University of Athens, Panepistimiopolis 157 84, Zografos, Greece}
\email{ahilaris@phys.uoa.gr}
}

\date{Received ...; accepted ...}

 
  \abstract
{We have studied a complex metric radio event which originated in a compact flare, observed with the ARTEMIS-JLS radiospectrograph on February 12, 2010. The event was associated with a surge observed at 195 and 304 \AA\ and with a coronal mass ejection observed by instruments on-board STEREO A and B near the East and West limbs respectively. On the disk the event was observed at 10 frequencies by the Nan\c cay Radioheliograph, in \ha\ by the Catania observatory, in soft x-rays by GOES SXI and Hinode XRT and in hard x-rays by RHESSI. We combined these data, together with MDI longitudinal magnetograms, to get as complete a picture of the event as possible. Our emphasis is on two type-II bursts that occurred near respective maxima in the GOES light curves. The first, associated with the main peak of the event, showed an impressive fundamental-harmonic structure, while the emission of the second consisted of three well-separated bands with superposed pulsations. Using  positional information for the type-IIs from the NRH and triangulation from STEREO A and B, we found that the type IIs were associated neither with the surge nor with the disruption of a nearby streamer, but rather with an EUV wave probably initiated by the surge. The fundamental-harmonic structure of the first type II showed a band split corresponding to a magnetic field strength of 18\,G, a frequency ratio of 1.95 and a delay of 0.23-0.65\,s of the fundamental with respect to the harmonic; moreover it became stationary shortly after its start and then drifted again. The pulsations superposed on the second type II were broadband and had started before the burst. In addition, we detected another pulsating source, also before the second type II, polarized in the opposite sense; the pulsations in the two sources were out of phase and hence hardly detectable in the dynamic spectrum. The pulsations had a measurable reverse frequency drift of about 2\,s$^{-1}$. } 


   \maketitle
%

\section{Introduction}

Type II bursts are narrow-band lanes of transient solar radio emission that appear to drift slowly toward lower frequencies over time in dynamic spectra (for detailed discussions about their properties, see the reviews by  \citeads{1985srph.book..333N}; \citeads{2008SoPh..253..215V}; \citeads{2008SoPh..253....3N}; \citeads{2008A&ARv..16....1P}). The instantaneous bandwidth of type II bursts may be as narrow as a  few MHz \citepads[\emph{e.g.}][]{1985srph.book..333N}.They typically start below 150 MHz \citepads[\emph{e.g.}][]{1996A&AS..119..489M} although cases with starting frequencies equal or higher than 500 MHz have been reported (\emph{e.g.} \citeads{2008A&A...490..357P}; \citeads{2012ApJ...746..152M}; \citeads{2013ApJ...765..148C}). In the meter wavelength range type II bursts last from less than 5 minutes to about 30 minutes and their drift rate lies between \mbox{0.1 to 0.4 MHz s$^{-1}$} and increases with increasing starting frequency (e.g. \citeauthor{1995A&A...295..775M} \citeyearads{1995A&A...295..775M}, \citeyearads{1996A&AS..119..489M}). These values of drift rates are consistent with MHD shocks propagating upwards in the corona and driving radio emission near the plasma frequency and/or its harmonic via the plasma emission mechanism.

Only about 60\% of coronal type II bursts display fundamental-harmonic emission  bands \citepads{1985srph.book..333N} and the fundamental band is usually weaker than  the harmonic band. It appears likely that the fundamental  band is absorbed, especially when the radio emission occurs behind the limb and passes through dense regions of the corona. Occasionally the fundamental and/or harmonic bands are divided into substructures, which are called multiple lanes when they show irregularly varying frequency ratios, and split bands when they show an  essentially constant frequency ratio, $\Delta f/f \approx 0.1-0.2$.  Multiple-lane events are interpreted in terms of emission from distinct source regions on the shock \citepads[\emph{e.g.}][]{2006pre6.conf..419C}. that can have different plasma densities, propagation speeds, and geometry allowing the emissions to appear separate and drift at different rates (\emph{e.g.} \citeads{1982PASAu...4..392R}, \citeads{2006A&A...448..739V}, \citeads{2015AdSpR..56.2811Z}).  A similar interpretation has been proposed for the split-band effect (\citeads{1967PASAu...1...47M}, \citeads{2012JGRA..117.4106S}). Alternatively, sometimes the band-splitting is interpreted in terms of plasma emission from both upstream and downstream of the shock \citepads{1974IAUS...57..389S} with the  frequency difference allowing the Alfv\'en Mach number of the shock to be determined (\emph{e.g.} \citeauthor{2001A&A...377..321V}, \citeyearads{2001A&A...377..321V}, \citeyearads{2002A&A...396..673V}, \citeyearads{2004A&A...413..753V}; \citeads{2014SoPh..289.2123K}). Although, there is no accepted model for plasma emission from the  downstream region \citepads[\emph{e.g.}][]{2011sswh.book..267C} observational support for this interpretation has been provided by \citetads{2012A&A...547A...6Z}, \citetads{2014ApJ...795...68Z} and \citetads{2018ApJ...868...79C}. 

Although it is generally accepted that interplanetary (IP) type II bursts are produced by coronal mass ejection (CME) driven shocks \citepads[\emph{e.g.}][]{2006GMS...165..207G}, there is no consensus about the origin of coronal type II bursts.  Coronal shocks could be generated by different drivers: CME-related  erupting structures, the pressure pulse of a flare, or flare-related  small-scale ejecta. Both flare-related drivers and CMEs could act as pistons with their main difference being the time over which energy is supplied by the piston  \citepads{2008SoPh..253..215V}. In the former case the piston is temporary and the shock freely propagates as a large-amplitude wave  (blast shock) whereas in the latter case the wave is supplied by the energy provided by the CME.

Statistical studies (see \citeads{2011A&A...531A..31N} and references therein) show  that most, if not all, coronal type II bursts are observed during events  with both flares and CMEs. Observations of metric type II bursts without  CMEs are rare and are usually attributed to source regions within about 30\degr\ from central meridian \citepads[\emph{e.g.}][]{2002A&A...384.1098C} where the detection of CMEs is more difficult. On the other hand, the lack of association of a metric type II burst with a flare is usually interpreted in terms of the occurrence of the flare behind the limb. In some events, there is a tight  synchronization between the flare impulsive phase and CME acceleration phase  (\emph{e.g.} \citeauthor{2001ApJ...559..452Z} \citeyearads{2001ApJ...559..452Z}, \citeyearads{2004ApJ...604..420Z}; \citeads{2008ApJ...673L..95T}) which makes the  identification of the shock driver difficult.  

Support of the CME-driven scenario has been provided by the high correlation between metric type II bursts and EUV waves (\citeads{2000A&AS..141..357K}; \citeads{2002ApJ...569.1009B}) although each one may appear separately \citepads[\emph{e.g.}][]{2014SoPh..289.4589N}. EUV waves (see the reviews by \citeads{2012SoPh..281..187P}; \citeads{2015LRSP...12....3W}; \citeads{2017SoPh..292....7L}) are thought to be driven by the lateral expansion of CMEs. Furthermore, \citeauthor{2005JGRA..11012S07G} (\citeyearads{2005JGRA..11012S07G}, \citeyearads{2009SoPh..259..227G}) proposed that  all metric type II bursts are produced by CME-driven shock waves by pointing out  that the Alfv\'en speed profile may exhibit a local maximum in the corona which could explain why metric type II bursts usually cannot be followed through longer wavelengths even if they are produced by the same CME. However, \citetads{2005ApJ...623.1180C} have argued against a common CME driver for both coronal and IP shocks.

Type II bursts limited to relatively high frequencies might be attributed to significant decay of the shock strength with height. Therefore they are usually associated to shocks driven by flare blast waves (\emph{e.g.} \citeads{1995SoPh..158..331V}; \citeads{2011A&A...531A..31N}; \citeads{2012ApJ...746..152M}) although exceptions have also been reported (\emph{e.g.} see \citeads{2011A&A...531A..31N} and \citeads{2017ApJ...843...10K} for high-frequency type II bursts associated with CMEs).

There are relatively few multi-wavelength case studies of the origin of metric type II bursts because (1) the metric emission comes from heights beneath the occulting disk of coronagraphs and (2) there are few interferometers that can provide routine observations of the Sun at metric wavelengths. Old Clark Lake and Culgoora imaging observations (see  \citeads{1983ApJ...268..403G}; \citeads{1984A&A...134..222G}, and also the review by \citeads{1985srph.book..333N}) show that in most cases the radio emission is located close to  the CME front or its flanks. 

More recent radio imaging  observations have strengthened and refined the above conclusion because they have been accompanied by dynamic spectra with high temporal and spectral resolution, high-quality coronagraphic images, and imaging data in the EUV and soft X-rays (SXR). Probably the largest recent compilation of imaging type II observations has been reported by \citetads{2012ApJ...752..107R} who found that 38 of the 41 type II sources that they observed at 109 MHz were associated with CMEs and were  located near their leading edge. A similar conclusion for five events observed in the 164-435 MHz range was reached by \citetads{2000ApJ...528L..49M}. The coronal shock radio sources may occur in front of an erupting flux rope (\citeads{2012ApJ...750...44B}; \citeads{2012A&A...547A...6Z}; \citeads{2015AdSpR..56.2811Z}), above expanding soft X-ray loops (\emph{e.g.} \citeads{1999A&A...346L..53K}; \citeads{2006A&A...455..339D}), in association with an erupting jet during the  progression of a CME (\emph{e.g.} \citeads{2014ApJ...795...68Z}; \citeads{2021ApJ...909....2M}), in front of an EUV bubble in both radial and lateral directions \citepads{2014SoPh..289.2123K} or ahead of a CME  that was deflected in the low corona \citepads{2016ApJ...823....5P}. Other studies indicate that type II sources are associated with the flanks of CMEs (\emph{e.g.} \citeads{2007A&A...461.1121C}; \citeads{2012ApJ...750..147D}; \citeauthor{2014A&A...564A..47Z}, \citeyearads{2014A&A...564A..47Z}, \citeyearads{2018A&A...615A..89Z}; \citeads{2019NatAs...3..452M}) CME-streamer interactions in the low corona could also be important for the production of type II bursts (\emph{e.g.} \citeads{2012ApJ...750..158K}; \citeads{2012ApJ...753...21F}; \citeads{2015AdSpR..56.2793E}; \citeads{2019A&A...624L...2M}; \citeads{2020ApJ...893..115C}).

In a smaller number of events the type II sources appear to be flare-related (\emph{e.g.} \citeads{white2007}; \citeauthor{2008SoPh..253..305M}, \citeyearads{2008SoPh..253..305M}, \citeyearads{2010ApJ...718..266M}, \citeyearads{2012ApJ...746..152M}; \citeads{2011A&A...531A..31N};  \citeads{2016ApJ...828...28K}). In these cases the type II events were tightly synchronized with the related flares and either no CME was observed or the CME was not synchronized with the type II burst. The above discussion shows that type II bursts may occur under significantly different conditions.

\begin{figure*}[t]
\centering
\includegraphics[width=.65\hsize]{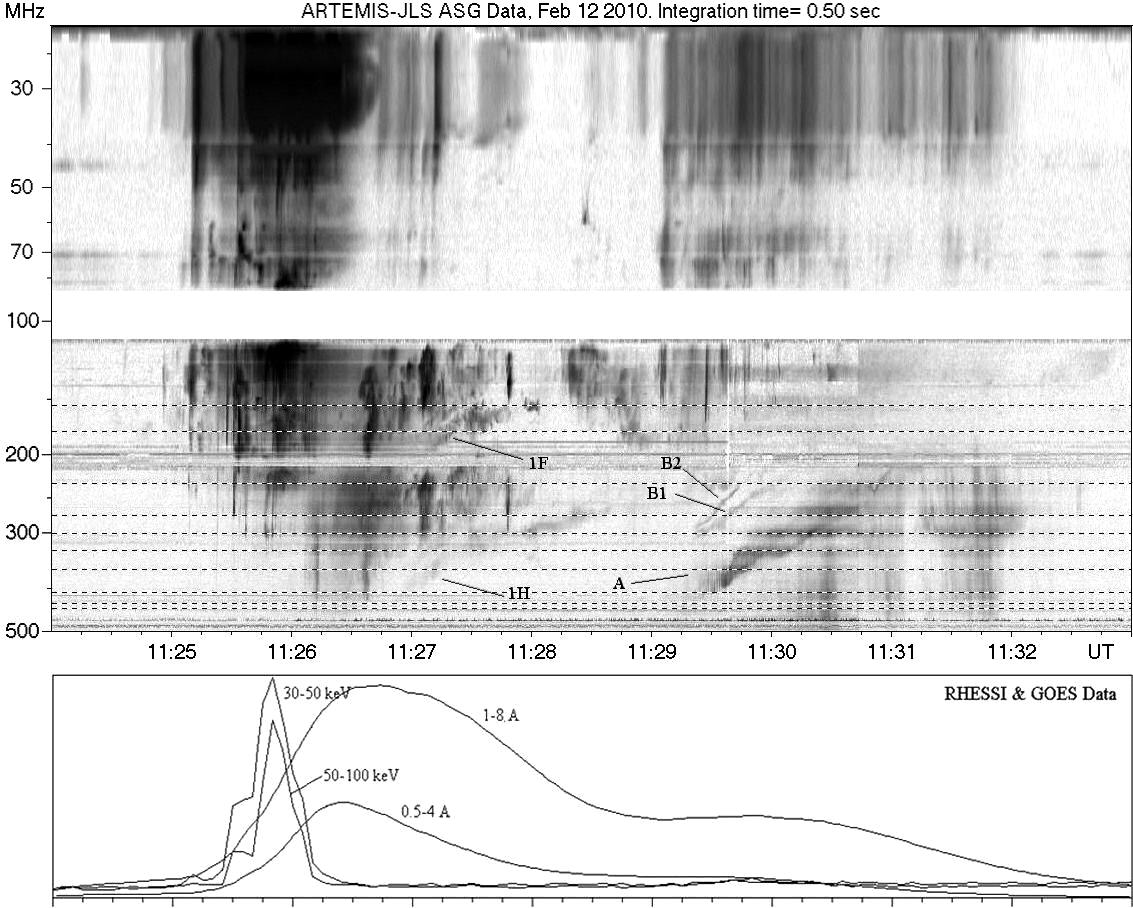}
\caption{Dynamic Spectrum (negative) of the February 12, 2010 event (500 to 20 MHz), observed with the ASG receiver of the ARTEMIS-JLS radiospectrograph, together with light curves in soft X-rays from GOES and hard X-rays from RHESSI in arbitrary units and on a linear scale. Dotted horizontal lines mark the frequencies of the Nan\c cay Radioheliograph. 1F and 1H mark the fundamental and harmonic emission of  the first type II burst, A, B1 and B2 mark the branches of the second type II.}
\label{figure1}
\end{figure*}
\begin{figure*}
\centering
\includegraphics[width=\hsize]{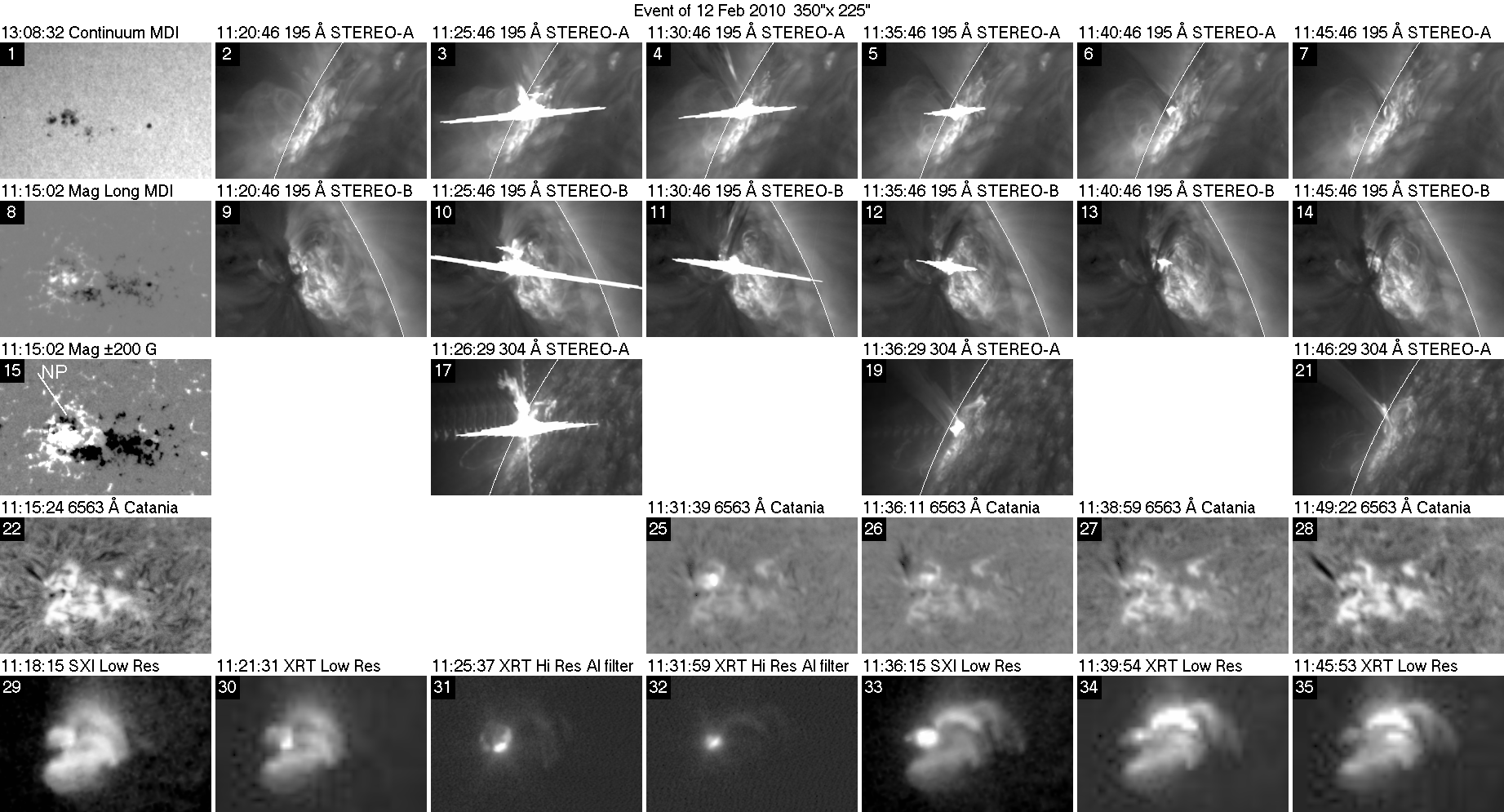}
\caption{Evolution of the event in \ha\ (Catania Observatory), soft X-rays (GOES/SXI and Hinode/XRT), 195 and 304 \AA\ (STEREO A \& B), together with a magnetogram and white light image from MDI. The white arc in the STEREO images marks the solar limb. The center of the field of view is at Carrington longitude 186.6, latitude 25.0 (solar $x=-107\arcsec, y=510\arcsec$  for Earth images). }
\label{figure2}
\end{figure*}

In this article we study a complex event, SOL2010-02-012T11:25:00, whose dynamic spectrum contained at least five lanes of type II emission and for which radio imaging data at 10 frequencies were available. It is thus one of the most complex type II events with adequate radio imaging data ever presented in the literature. We combine the radio spectral and imaging data with coronagraphic images and EUV and soft X-ray images from a variety of sources including {\it Solar Terrestrial Relations Observatory} (STEREO) Ahead (A) and Behind (B), in an attempt to associate the type-II-related shocks with disturbances in the corona. We present our data in Sect.~\ref{obs} and our results in Sect.~\ref{Results}. In Sect.~\ref{Discuss} we discuss the origin of the type II bursts and in Sect.~\ref{Concl} we summarize our conclusions.

\section{Observations}\label{obs}
Observations with the {\it Appareil de Routine pour le Traitement et l' Enregistrement Magnetique de l' Information Spectral-Jean Louis Steinberg} (ARTEMIS-JLS) radiospectrograph \citepads{2006ExA....21...41K} and their analysis have 
been described in detail in previous articles (\emph{e.g. } \citeads{2011A&A...531A..31N}; \citeads{2015SoPh..290..219B}; \citeads{2019A&A...627A.133A}). Here we only mention that the event was observed in high temporal resolution (10\,ms) with the {\it acousto-optic analyzer} (Spectrograph Acousto-Optic, SAO) of the instrument in the frequency range of 265--470\,MHz with a spectral sampling of 1.4\,MHz. For an overview of the event we used data from the sweep-frequency {\it Global Spectral Analyser} (Analyseur de Spectre Global, ASG), which operated in the 100-700\,MHz range with a time resolution of 100\,ms and a spectral sampling of 1\,MHz.

The Nan\c cay Radioheliograph (NRH; \citeads{Kerdraon97}) is a synthesis instrument that provides 2D images of the Sun with sub-second time resolution. For the event that we study here, the NRH provided data at ten frequencies (150.9, 173.2, 228.0, 270.6, 298.7, 327.0, 360.8, 408.0, 432.0 and 445.5\,MHz) with a cadence of 250 ms. All ten frequencies were within the spectral range of the ASG, while the last eight were also within the range of the SAO. 

The observed source position may oscillate due to ionospheric effects, which are stronger in  the winter and for low solar elevation. Indeed, such oscillations affected the position of a noise storm continuum visible in the full-day summary NRH plots \footnote{\url{http://secchirh.obspm.fr/spip.php?page=survey&hour=day&survey_type=1&dayofyear=20100212}}. Our measurements showed that the rms amplitude of the ionospheric oscillations at 150.9\,MHz was 74\arcsec\ at the start of the daily NRH observations but dropped to 30\arcsec\ before the start of our event, which corresponds about 7\% of the average source size (about 430\arcsec) at the same frequency. The shift was even smaller at high frequencies (6\arcsec, or about 3\% of the source size at 408.0\,MHz), thus ionospheric effects could be safely ignored.
 
From the original NRH visibilities, we computed 2D images with a resolution of 1.2\arcmin\ by 1.9\arcmin\ at 432 MHz. We also computed 1D images using visibilities from the east-west (EW) and north-south (NS) arrays only; this improved the resolution by a factor of two, due to the fact that the NRH extension antennas have a very small contribution to the 2D images \citepads[\emph{c.f.}][]{2016A&A...586A..29B}. 

In addition to the ARTEMIS-JLS and NRH data we used light curves from the Geostationary Operational Environmental Satellite (GOES) and the Reuven Ramaty High Energy Solar Spectroscopic Imager (RHESSI), \ha\ images from Catania Observatory, soft X-ray images from the GOES {\it Soft X-ray Imager} (SXI) and the Hinode {\it Soft X-ray Telescope} (XRT), EUV images (195 and 304 \AA) from STEREO {\it Sun-Earth Connection Coronal and Heliospheric Investigation} (SECCHI) Ahead and Behind \citepads{2008SSRv..136...67H}, coronagraph images from STEREO {\it Coronagraph1} (COR1) A and B \citepads{2008SSRv..136...67H}, as well as white light images and magnetograms from the {\it Michelson Doppler Imager}, (MDI) before and after the event. No Solar and Heliospheric Observatory (SOHO) {\it Extreme ultraviolet Imaging Telescope} (EIT) data were available during the event, while the {\it Transition Region and Coronal Explorer} (TRACE) was pointed elsewhere.

We only had four \ha\ images and those after the start of the event, with one of them during the radio burst. The SXI provided 15 images during the event (9 during the burst), with various  exposure times and filters; the pixel size was 5\arcsec. From the XRT we obtained a set of low resolution images with a pixel size of 8.2\arcsec\ and a Ti poly filter covering the entire event with a cadence of 2\,min and two sets of high resolution images (pixel size of 1\arcsec) with Al thick and Be thick filters with a 20\,s cadence, starting in the early phase of the event. Among the SECCHI images, those from STEREO-A in the 195\,\AA\ band had the highest cadence, 150\,s; the cadence from STEREO-B was 5\,min. In the 304\,\AA\ band the cadence was 10\,min, giving only one image during the metric burst. In the 171 and 284\,\AA\ SECCHI bands the event was not recorded due to the very low image cadence (2\,h). We note that several EUV and soft X-ray images suffered from saturation effects.

COR1 gave images with a cadence of 5\,min, with the first CME image beyond the occulting disk, located at 1.6\,R$_\odot$, having been after the radio burst. We did not use any {\it Coronagraph2} (COR2) images because the CME passed beyond the occulting disk well after the burst. 

\section{Results}\label{Results}
\subsection{Overview of the event}\label{Overview}
The event studied in this work was the third and strongest of a set of 3 homologous events, described by \citetads{2014Ge&Ae..54..406C}; it occurred around 11:25 UT on February 12, 2010 in active region 11046. It was a fairly strong (GOES class M9) and complex event, with an associated CME, the main phases of which are given in Table~\ref{tab:timeline}. Preliminary results were reported by \citet{Alissandrakis2011}. The other two events on that day were weaker than the event studied here; they occurred around 07:21 UT (C7.9) and 09:40 UT (B9.6). Their dynamic spectra were rich in fine structure, see \citetads{2014Ge&Ae..54..406C}, but did not show any type II emission. The 07:21 UT event was accompanied by a CME. We add that a noise storm continuum associated to the active region was observed by the NRH from February 10 to February 17, 2010.

Fig.~\ref{figure1} shows the Dynamic Spectrum (DS) of the event, recorded with the ASG (sweep frequency) receiver of the ARTEMIS-JLS radio spectrograph, together with the GOES and RHESSI light curves. The DS shows that the event was rich in fast drift, type-III, U and other structures. Both the light curves and the spectrum indicate two phases, the first significantly stronger than the second. The second phase started at $\sim$ 11:28:15 UT with type III bursts in the low frequency part of the dynamic spectrum. A secondary maximum appeared around 11:30 in the GOES and RHESSI light curves. The bulk of the radio emission ended around 11:32 UT, but metric continuum emission appeared from 12:00 to 12:45 UT (see Fig. 45 of \citeads{2015SoPh..290..219B}); this emission will not be considered in this work because it had no associated type II bursts.

Two type-II bursts appeared, one in each phase of the event, in the metric part of the spectrum (>100 MHz). The first type II showed fundamental-harmonic (FH) structure, marked 1F and 1H in Fig. \ref{figure1}, as well as band splitting (see Sect.~\ref{fh} for more details); no herringbone structures were detected. The second type II differed from the first in that it had three branches with different drift rates, marked A, B1 and B2 in the figure, with B1 and B2 starting 20\,s after A, with superposed pulsations (see also Sect.~\ref{puls}); there was no FH structure, no band splitting and no herringbones. At even lower frequencies, WIND/WAVES and STEREO/WAVES data showed no extension of the type IIs, but only two strong type-III bursts, which were extensions of the type-III groups that appeared in the high frequency of the spectrum at the beginning of each phase.

\begin{figure*}
\centering
\includegraphics[width=.95\textwidth]{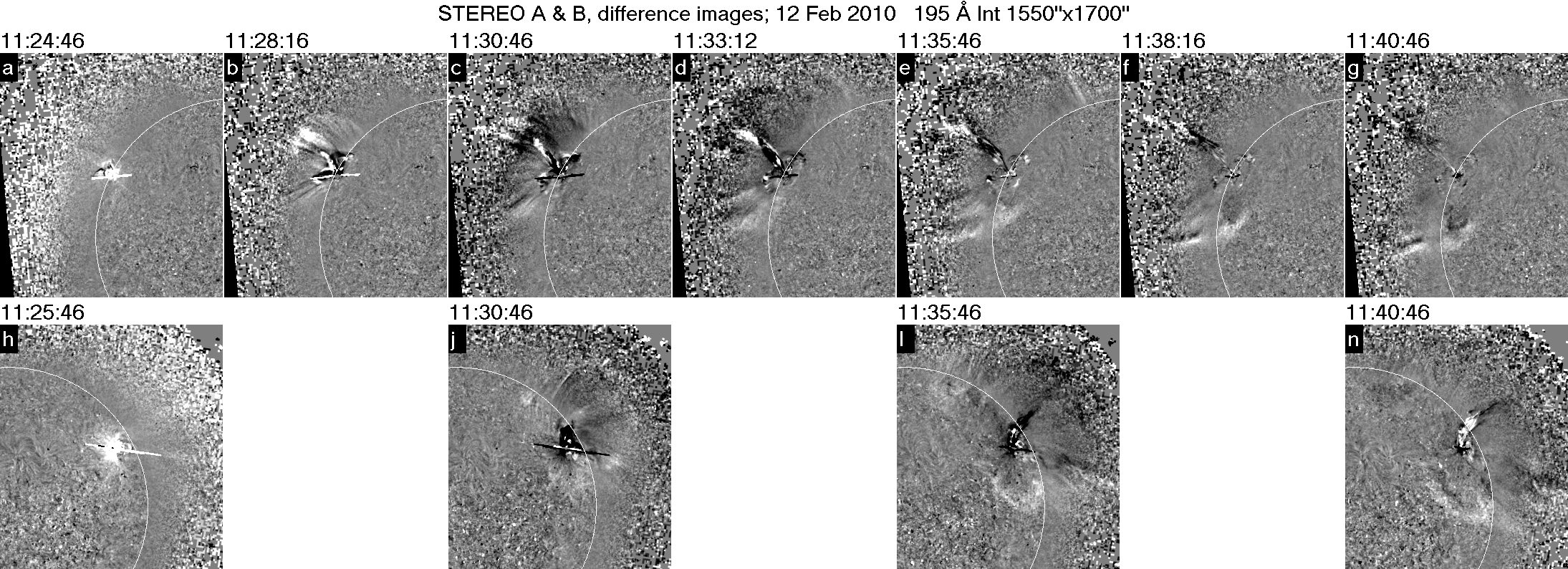}
\caption{Running difference images at 195 \AA\ from STEREO A \& B. Note the formation of an expanding EUV wave. The white arc marks the limb.}
\label{figure3}
\end{figure*}

Seen from the earth, the event occurred near the central meridian (N25, E11; heliocentric distance 33\degr). Its evolution, through images from SXI, XRT, SECCHI (STEREO A and B), \ha, as well as white light images and magnetograms is presented in Fig.~\ref{figure2}. From the position of STEREO A the flare was very close to the East limb, while from STEREO B it was seen near the West limb. 

\begin{table}
\begin{center}
\caption{Time line of the event of 12 February 2010 }
\begin{tabular}{lcl}
\hline
Time &Wave &Event\\
\hline
11:20:46&EUV&   First brightening detected in 195 \AA\ images\\
11:23     &EUV&   First evidence of surge in 195\,\AA\ images\\
11:24:15&Metric& First type III in decametric-$\lambda$\\
11:25     &Metric& Start of first type III storm\\
             &EUV&   Estimated maximum acceleration of surge\\
11:25:46&EUV&   First signature of EUV wave in 195\,\AA\ image\\
11:25:50&HXR&   RHESSI peak\\
11:26:30&SXR&   GOES 0.5-4\,\AA\ peak\\
             &WL& Extrapolated CME start\\
11:26:40&SXR& GOES 1-8\,\AA\ peak\\
11:26:50&Metric& Start of first type II burst\\ 
11:28:15& All & Start of second phase of the event\\
             &Metric& Precursor metric type IIIs\\
11:28:16&EUV&   Disruption of N loops in 195\,\AA\ image\\
11:28:40&Metric& End of first type II burst\\ 
11:29     &Metric& Second type III storm\\
11:29:05&Metric& Start of second type II\\
11:29:50&SXR&   GOES 1-8\,\AA\ second peak\\
11:32     &Metric& End of the bulk of metric radio emission\\
11:35:21&WL&   Streamer disruption detected in COR1-A\\
13:26     &EUV&   Surge still visible in SECCHI 304\,\AA\ images\\
\hline
\label{tab:timeline}
\end{tabular}
\end{center}
\end{table}

The left column of Fig.~\ref{figure2} shows the pre-flare situation, where a set of loops are visible in the SXR image (panel 29); the magnetogram is shown twice, saturated at $\pm200$\,G in panel 15. The first brightening appeared near 11:20:46 UT in the 195\,\AA\ STEREO-A image; it is well visible in the XRT image at 11:21:31 (panel 30 in Fig.~\ref{figure2}), and was located near a compact negative polarity region adjacent to a positive polarity spot, marked as NP in panel 15 of the figure. We note that this happened about 3.5 min before the first type III occurred. 

The bright point evolved to a compact source; in the XRT high resolution image (panel 31 in the figure) it appears elongated, with approximate full width at half maximum of about $32\arcsec\times15\arcsec$, probably a compact loop. A second, smaller compact source appeared at the beginning of the second phase of the event in high resolution SXR images, located $\sim8$\arcsec\ NE of the first (panel 32); by that time the peak of the previous bright point had shifted slightly to the north. In panels 33-35 a set of loops NW of the compact source is visible in the SXI images, with bright footpoints visible in \ha; these appeared during the second phase of the event, indicating a restructuring of the magnetic field.
We note that, apart from these bright points, no extended flare ribbons were observed either in SXR or \ha.

Around 11:23 UT, ejecta in the form of a surge appeared in the STEREO images.
The surge had both a hot component visible in emission at 195 \AA\ and a cool component visible in emission at 304 \AA\ and in absorption at 195 \AA\ and \ha\ (Fig.~\ref{figure2}); however, we could not identify any trace of the surge in the SXR disk images. From the \ha\ images (panels 25-27 in the figure) it is obvious that the surge did not originate in the compact source, but $\sim30\arcsec$ to its east: we note that surge activity in \ha\ was observed prior to and after the flare (panels 22 and 28 respectively), at different locations. In the SECCHI images the surge was visible well after the end of the metric burst, up to about 13:26 UT.

There was no flux rope proxy (\emph{e.g.}, hot channel, sigmoid, etc; see for example Table-1 in \citeads{2020SSRv..216..131P}) in the employed limb and disk  imaging EUV and SXR data. The surge  itself,  as observed in the Extreme Ultraviolet Imager (EUVI) 195 A of \mbox{STEREO-A} contained seemingly intertwined  bright and dark threads, in emission and absorption respectively,  which might be an indication of the flux rope structure; in EUVI 195 \AA \, of STEREO-B the surge appears almost exclusively in absorption. On the other hand, inspection of the 304 \AA \, movies in both STEREO  A and B, suggests that the ejected plasma was rather laminar.

Fig.~\ref{figure3} shows running difference images at 195\AA\ from STEREO A (top row) and  from STEREO B (bottom row). The eruption produced a disturbance expanding in the corona, in the form of an EUV wave 
\citepads[\emph{e.g.} ][]{1998GeoRL..25.2465T}. Essentially the limit of the EUV wave separates the coronal region already affected by the eruption from the region which has not been affected yet. Note that, as the disturbance expands, it is better visible in the south of the event than in the north in the STEREO-A images. An interesting remark is that the top of the surge appears to be close or even beyond the limit of the EUV wave (panels d and e of the figure). The trace of the base of the EUV wave is clearly seen on the disk in panels d-g (STEREO-A) and i-k (STEREO-B) in Fig.~\ref{figure3}). The EUV wave expanded outside the flaring active region and propagated over significant distances including  surrounding quiet Sun areas. In addition, during its propagation, it caused deflections of ambient structures, best observed along its northern track in STEREO-A. We note in particular a disrupted set of loops N of the surge, better visible in Fig.~\ref{195+typeII} below; these might be associated to the restructuring of loops seen in the SXR images during of the second phase of the event, as noted above.

\begin{figure}
\centering
\includegraphics[width=\hsize]{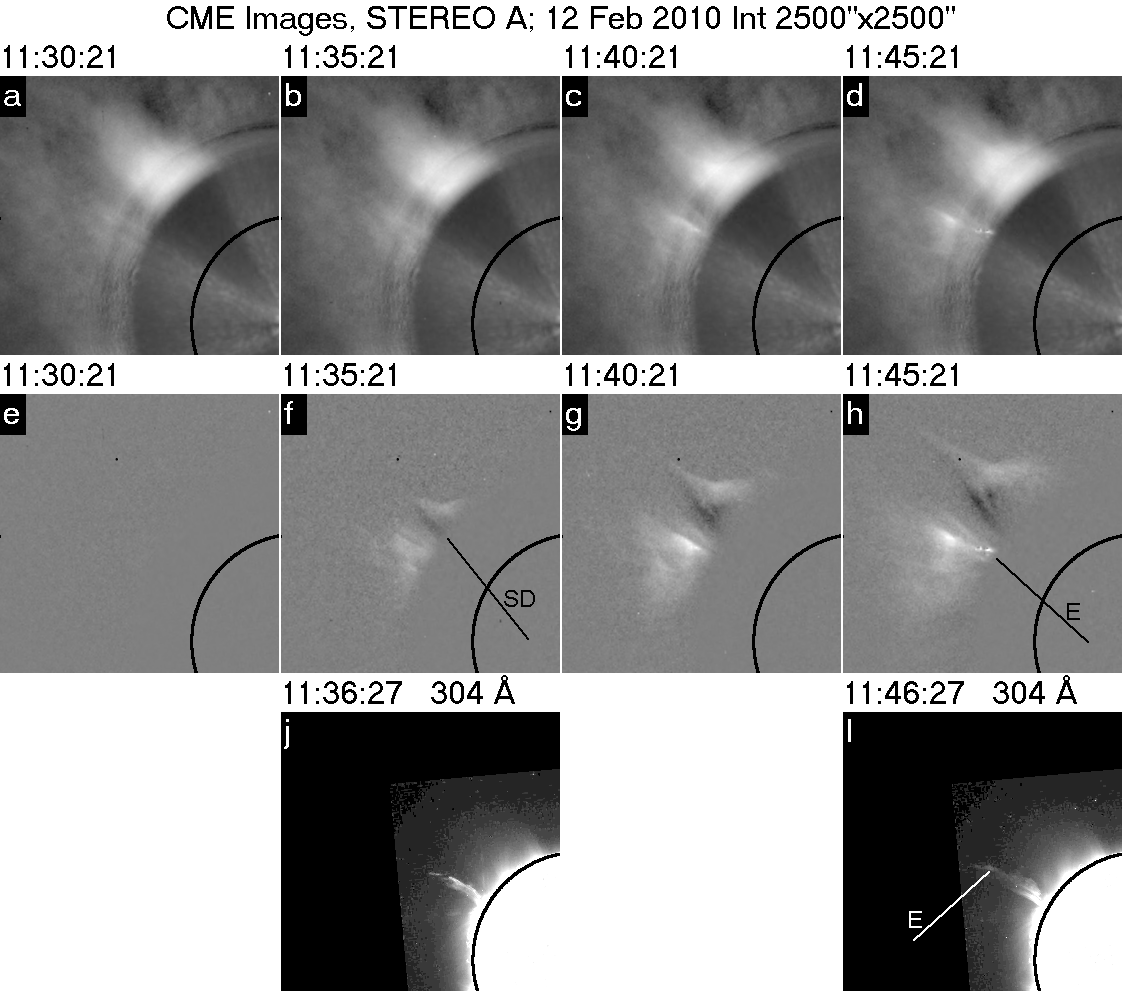}
\caption{COR1 images and their base difference (top and middle row) and 304 \AA\ images, saturated to show the surge better. Note the appearance of the CME above the occulting disk at 1.6 solar radii at 11:35:21 UT, the disruption of the streamer (marked SD) and the surge (E) seen both in the coronagraph and the 304 \AA\ images. The dark circle marks the limb. The field of view is 2500 by 2500\arcsec.}
\label{figure4}
\end{figure}

The white light images from the STEREO-A coronagraph COR1 (Fig.~\ref{figure4}, top row) show a streamer north of the active region that produced the flare; the CME was first seen above the occulting disk (1.6\,R$_\odot$) at 11:35:21 UT. The difference images (middle row of the figure), computed by subtraction of the average of images at 11:25:21 and 11:30:21 UT, clearly show a deflection of this streamer to the North by about 150\arcsec (marked SD on the figure), apparently a result of interaction with the expanding coronal disturbance. 

The first image where the streamer deflection was recorded above the coronagraph occulting disk was after the end of the metric event. With a 5 min cadence and the occulting disk at 1.6\,R$_\odot$ it is hard to specify when this deflection started and at which height, it is thus quite possible that the deflection started earlier at lower heights. Indeed, the 195\,\AA\ STEREO-A image at 11:28:12 UT shows a disruption of loops at the base of the streamer, well visible in Fig.~\ref{195+typeII} below at position angle of 40\degr. It appears that the streamer was an obstacle to the propagation of the disturbance to the north, and this is the reason that its trace was better visible south of the event in the 195\,\AA\ images, as noted in the previous paragraph. A final remark is that, in the last column of Fig.~\ref{figure4}, the surge is visible both in the coronagraph and the 304\,\AA\ images.

\subsection{Triangulation and velocity estimates}
The cadence of STEREO images is not sufficient to give accurate velocities from triangulation. However, we could measure the apparent motion of the top of the surge in 195 \AA\  STEREO-A images; we found that the radial component of the velocity was about 500\,km\,s$^{-1}$ and that the acceleration peaked around 11:25 UT (i.e. near the images shown in the first column of Fig.~\ref{figure3}). This roughly coincides with the start of the bulk of the type III emission, but it is 2 min before the start of the first type II. 

\begin{figure}[h]
\centering
\includegraphics[width=.7\hsize]{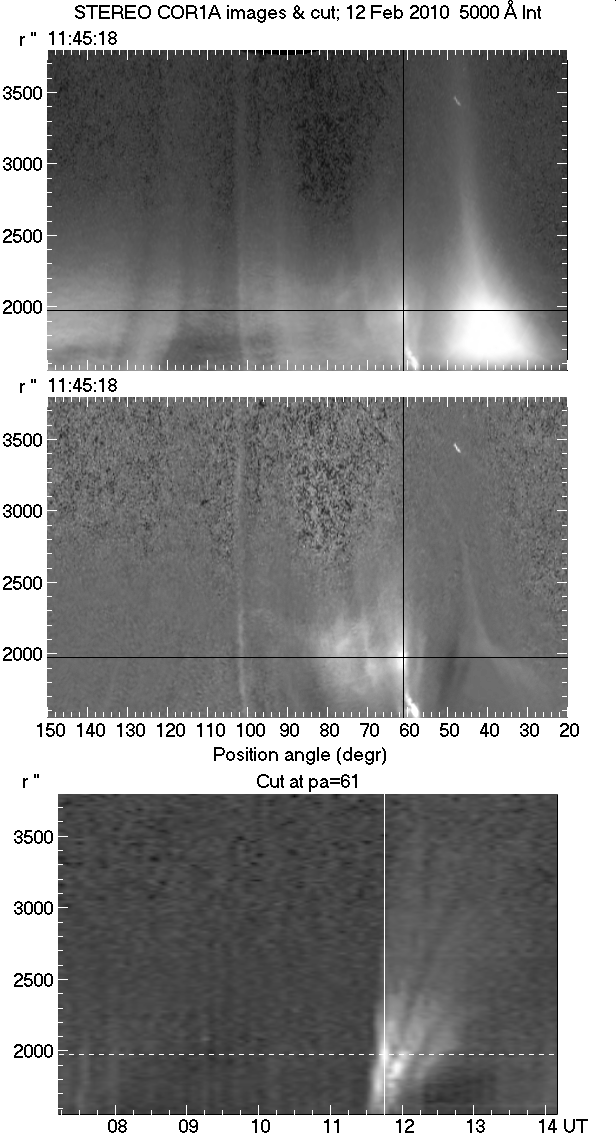}
\caption{Top: COR1A intensity as a function of position angle and distance from the center of the disk, $r$. Middle: Same as top, with the pre-CME intensity subtracted. Bottom: Cut of intensity as a function of time and $r$ at a position angle of 61\degr. Horizontal lines are at the same $r$ in all panels. Vertical lines in the top and middle panel mark the position angle of the cut. In the bottom panel the vertical line marks the time of the other panels. The photospheric limb is at $r=972$\arcsec.}
\label{cor1a+cut}
\end{figure}

In addition, we were able to apply triangulation on several points along the surge to the 195 \AA\ images at 11:30:46 UT; this gives a rough idea of the shape and, possibly, the trajectory of the surge. The maximum radial distance measured in this way was 1.27\,R$_\odot$, that is 265\arcsec\ above the photosphere.  By fitting a straight line to the positions of these points, we found that the projection of the surge on the photosphere was oriented 39\degr\ east of north, whereas its angle with respect to the vertical was 36\degr. 

It is also possible to estimate velocities by measuring positions in the original images, in particular those of STEREO-A where the flare appeared very close to the limb. To this end we computed cuts of intensity along the radial direction, as a function of position and time, for a number of position angles. An example is shown in Fig.\ref{cor1a+cut}, where a radial cut of COR1-A intensity is shown, together with associated images. 

Three features, moving at different radial velocities are visible in the cut: the fastest is a blob in the CME, also visible in panel d of Fig.~\ref{figure4},  moving at  $680\pm80$\,km\,s$^{-1}$, not too different from the speed of the 195\,\AA\ surge mentioned in the previous paragraph. A slower feature, at $244\pm16$\,km\,s$^{-1}$, corresponds to the 304\,\AA\ surge seen in the images of the right column in Fig.~\ref{figure4}; the lower speed of this feature, compared to the speed of the 195\,\AA\ surge, might be due to the fact that in the latter case the top of the surge was measured whereas here the average position of the blob, which apparently expanded radially, was recorded. The slowest feature, at $120\pm8.5$\,km\,s$^{-1}$, is associated to a second ejection seen in 304\,\AA\ images later during the event, visible at 60\degr\ position angle in the lower part of the middle panel of the figure. Velocity errors quoted here and later in the text correspond to the rms of values obtained at several position angles. 

A linear extrapolation of radial positions placed the start of the CME around 11:26:30 UT. There was a data gap in SOHO/LASCO C2, whereas C3 recorded a halo CME centered around a position angle of 44\degr\ and moving at 509\,km\,s$^{-1}$; the estimated onset time was 11:18:22 UT. Taking into account the geometry of the event, the present estimates from COR1 should be more accurate than those from C3.  

Applying the same technique to the CME associated to the earlier event of 07:21 UT (first CME), we detected a single feature moving at $\sim740$\,km\,s$^{-1}$ and no evidence of surge.

\begin{table}
\begin{center}
\caption{Measurements of velocities from WL and EUV images}
\begin{tabular}{lcc}
\hline
Feature& Direction & Velocity\\
           &              & km\,s$^{-1}$\\
\hline
Tip of 195\,\AA\ surge & Radial & $\sim500$ \\
White light CME, LASCO C3& Radial & 509 \\
White light CME, this work           & Radial & $680\pm80$ \\
White light surge 1      & Radial & $244\pm16$ \\
White light surge 2      & Radial & $120\pm8.5$ \\
White light CME           & Lateral & $\sim320$ \\
Expanding EUV wave, ST-A 195\,\AA\ &Radial&$460\pm10$\\
Expanding EUV wave, S, ST-A 195\,\AA\ &Lateral&$400\pm10$\\
Expanding EUV wave, N, ST-A 195\,\AA\ &Lateral&$560\pm22$\\
\hline
\label{tab:vel}
\end{tabular}
\end{center}
\end{table}

In a similar way we measured the lateral expansion of the southern front of the CME, using image cuts parallel to the limb. In the height range 1080-1530\arcsec\ above the limb we found velocities of $\sim320$\,km\,s$^{-1}$, less than half the value of the radial expansion. The corresponding value for the first CME was $410\pm20$\,km\,s$^{-1}$.

The southern part of the expanding EUV wave in the 195\,\AA\ band shown in Fig.~\ref{figure3} and discussed in the previous section is well visible in image cuts parallel to the limb; the northern edge of the EUV wave is also detectable, though much weaker. At the radial distance of the flare, we measured lateral expansion velocities of about 400\,km\,s$^{-1}$ and 560\,km\,s$^{-1}$ for the south and north fronts of the disturbance respectively.

The southern front of the expanding EUV wave can be followed above the limb, with its lateral velocity decreasing with $r$, down to $\sim350$\,km\,s$^{-1}$ at $r=1200$\arcsec, approaching the lateral velocity of the CME. This is apparently a geometric effect, for a roughly spherical expanding disturbance. 

Finally, the radial velocity of the EUV wave was measured from cuts in the radial direction, giving a value of about 460\,km\,s$^{-1}$. The velocities measured from white light (WL) and EUV images are summarized in Table~\ref{tab:vel}.

\subsection{Position of the type II sources and other features with respect to the surge}\label{positions}
\begin{figure}
\centering
\includegraphics[width=\hsize]{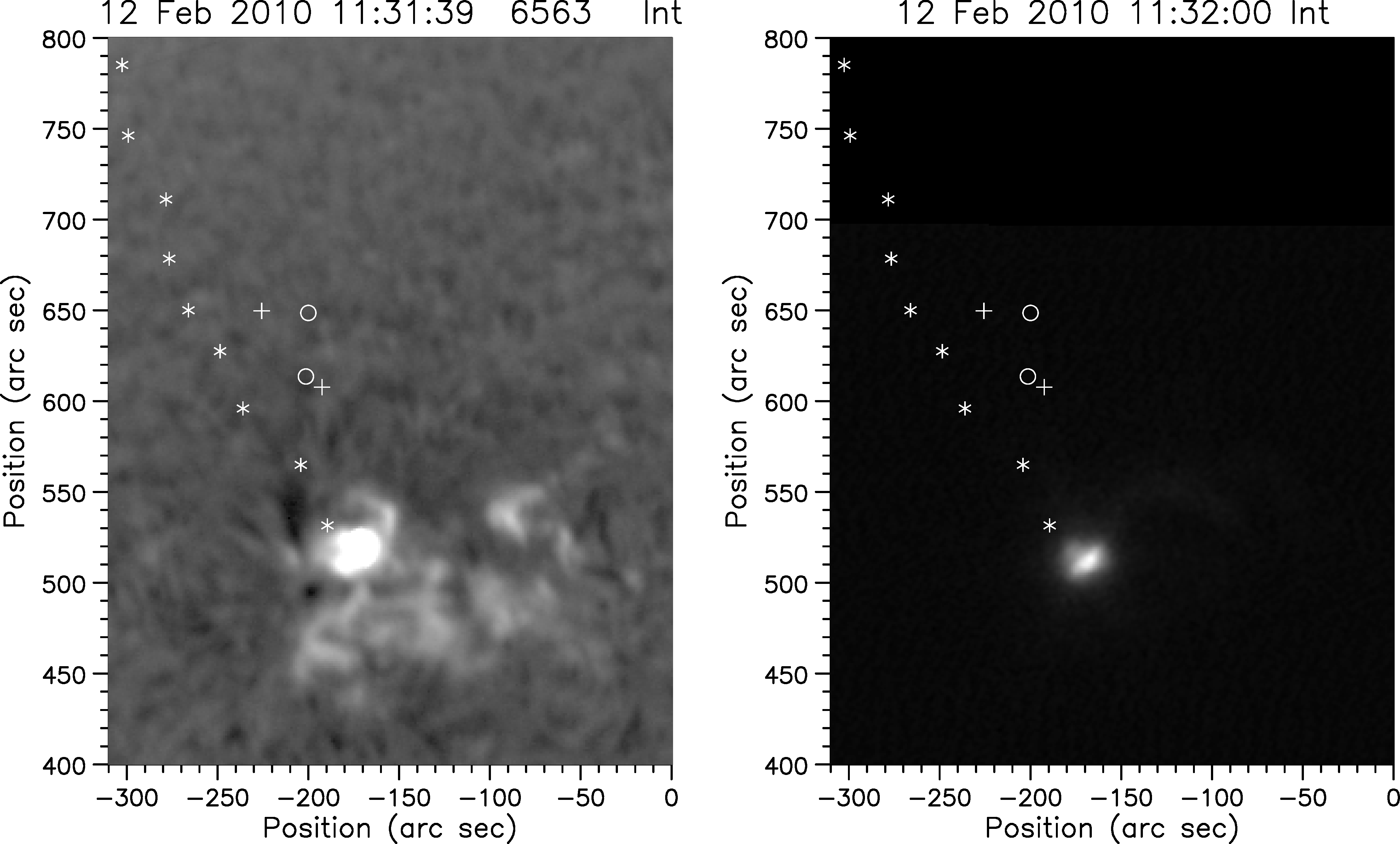}
\caption{Projection of the surge on the sky plane as seen from the Earth (symbols), on top of an \ha\ image (left) and an XRT image (right). Different symbols refer to different parts of the surge, giving an idea of its lateral extent.}
\label{Triang+Halpha+SXR}
\end{figure}

In this section we will compare the information on the surge obtained by triangulation with the positions of the type II bursts and other features seen from the earth. Fig.~\ref{Triang+Halpha+SXR} shows the projection of the surge position on top of \ha\ and SXR images. We note that the extrapolation of the surge positions to lower heights passes very close to the \ha\ and SXR flares; we also note that the \ha\ surge is located at a different position and has a slightly different orientation, showing that it is not part of the same phenomenon. 

\begin{figure}
\centering
\includegraphics[height=6cm]{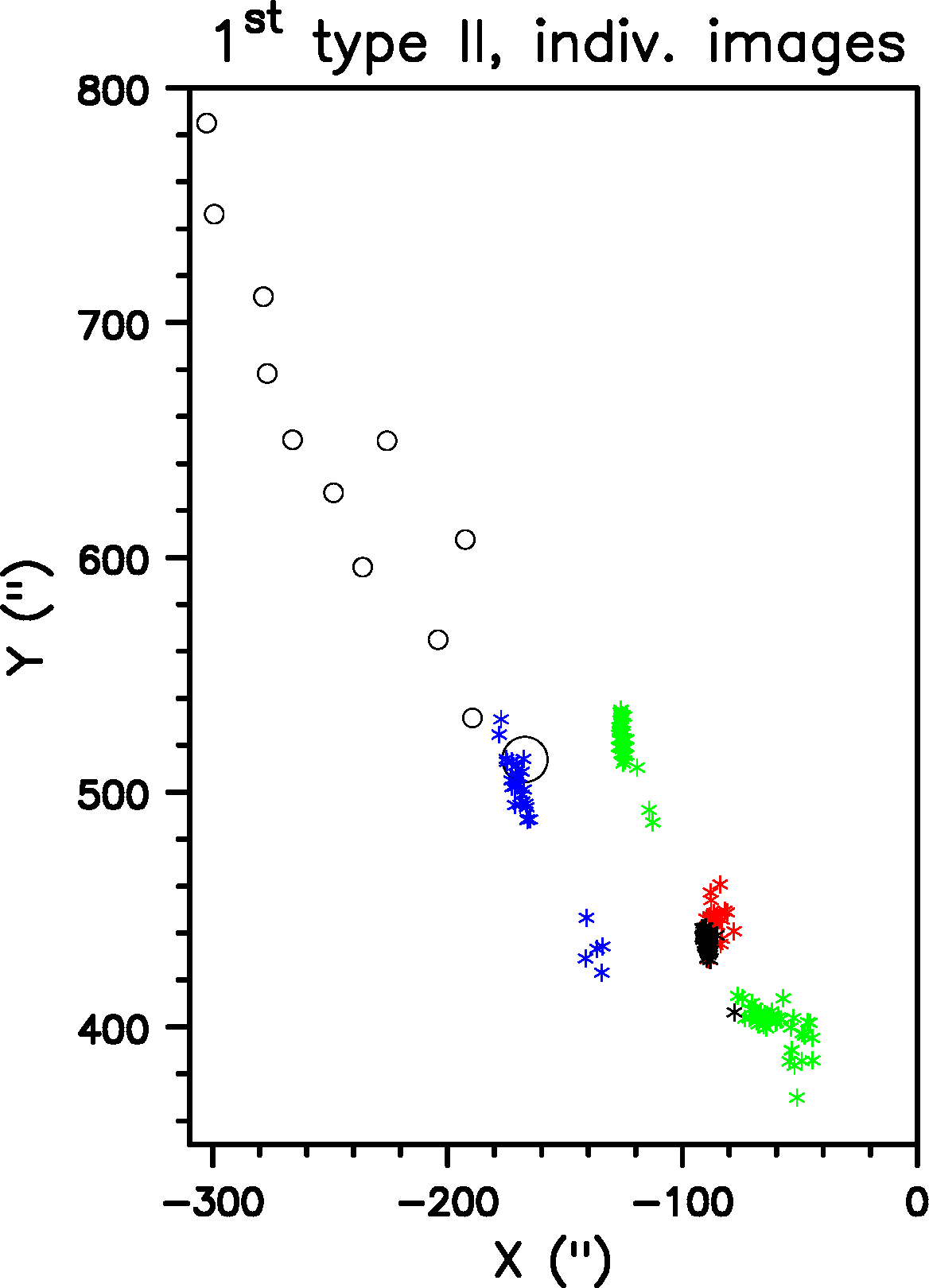}~\includegraphics[height=6cm]{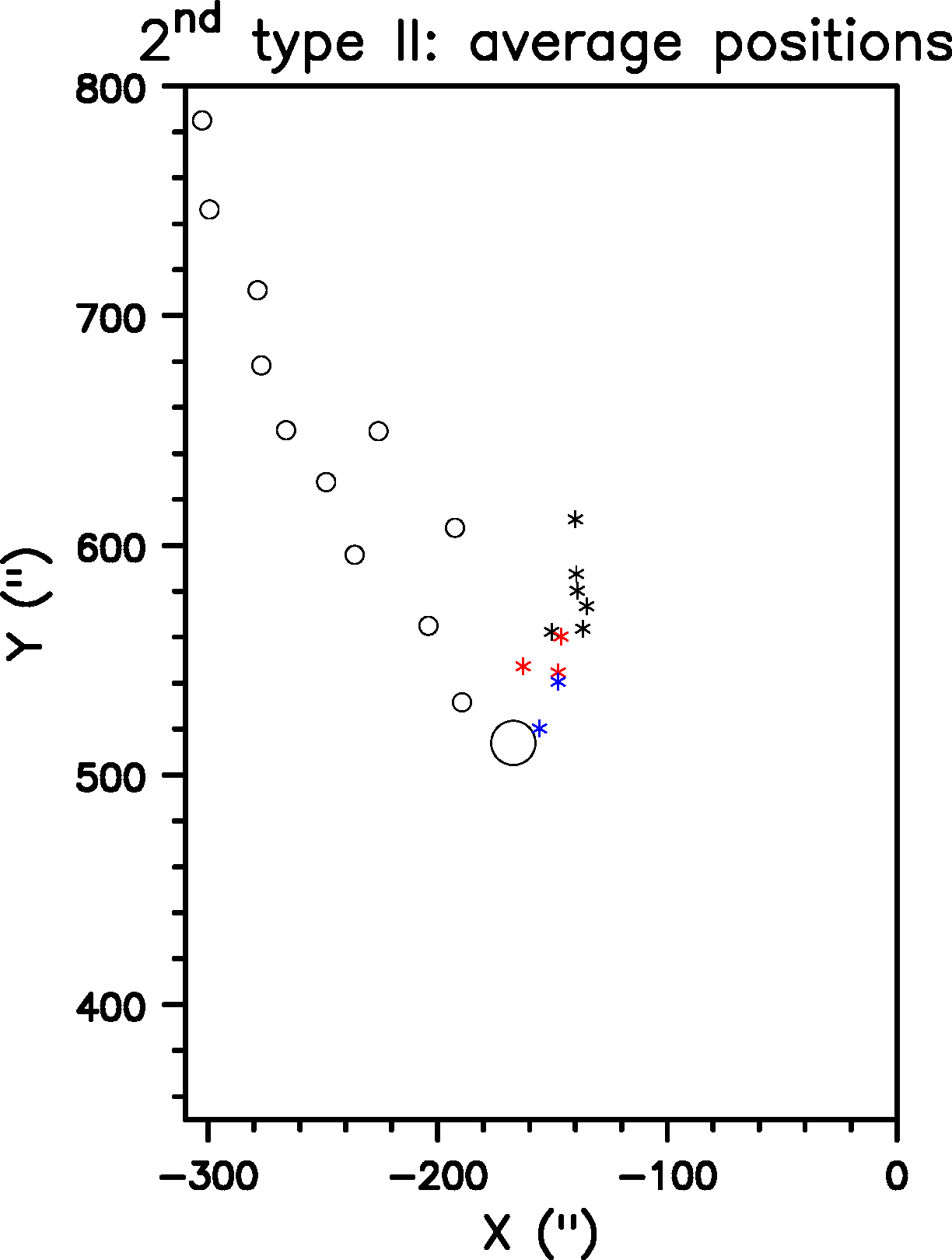}
\caption{Position of the type II sources, together with the projected position of the surge (open circles) and the position of the \ha/SXR flare (large open circle). Left: First type II, individual images; black is for 360.0\,MHz, red for 327.0\,MHz, green for 297.8\,MHz and blue for 150.9\,MHz. Right: Average images for branches A (black), B1 (red) and B2 (blue) of the second type II burst.}
\label{typeII+jet}
\end{figure}

Fig.~\ref{typeII+jet} shows the positions of the NRH sources during the first and second type II bursts. As the DS is rich in features, during the first phase of the event in particular, we took care to avoid emissions which were not related to the type IIs. Before proceeding we should note that the observed positions may deviate from the true positions due to refraction and scattering effects; for a spherically symmetric corona, refraction moves the source closer to the disk center, whereas scattering (\citeads{1971A&A....10..362S}; \citeads{1977A&A....61..777B}; \citeads{1994ApJ...426..774B}) moves it in the opposite direction and, at the same time, broadens the source. These effects are stronger for emission at the fundamental, where the refraction index deviates significantly from unity. 

In recent works (\emph{e.g.} \citeads{2019ApJ...884..122K}; \citeads{2021ApJ...909..195Z} and references therein) both refraction and anisotropic scattering are considered. \citetads{2021ApJ...909..195Z} employed a spherically symmetric coronal model at 35\,MHz and, for the heliocentric position of our flare, gave radial shifts of 0.08 to 0.24\,R$_\odot$ for the fundamental and $-0.05$ to 0.03\,R$_\odot$ for the harmonic, negative values denoting shifts towards the center of the disk (their Figure 7). According to \citetads{2019ApJ...884..122K}, the width and the decay time of the source scale approximately as $f^{-1}$; if this scaling also holds for the shifts, they should be much smaller than the above values for our frequency range, a factor of 4 to 13 higher than 35\,MHz, and probably negligible for the harmonic. Indeed, scattering computations (Kontar 2021, private communication) showed that the source displacement for the harmonic at 300\,MHz is 7\arcsec\ and 1.1\arcmin\ for the fundamental at 150\,MHz, both considerably smaller than the NRH beam size (1.5\arcmin and 4.4\arcmin\ respectively); hence no corrections were applied.

The most compact group of points in the left panel of Fig.~\ref{typeII+jet} is that from 360\,MHz; this is located about 80\arcsec\ away from the flare and further away from the surge. Although the radio sources are big, from about 150\arcsec\ by 90\arcsec\ at 408.0\,MHz to 400\arcsec\ by 260\arcsec\ at 150.9\,MHz, their positions can be measured with much higher accuracy; hence the difference with respect to the surge position is real.  Positions at 150.9\,MHz show the largest scatter. In general the radio emission does not appear to be associated with the surge, although the radio sources are aligned roughly in the same direction. For the second type II, the average position of the radio sources are at a different location, close to the flare and the lower part of the surge, but they are aligned along a direction quite different from that of the surge.

\begin{figure}
\centering
\includegraphics[width=\hsize]{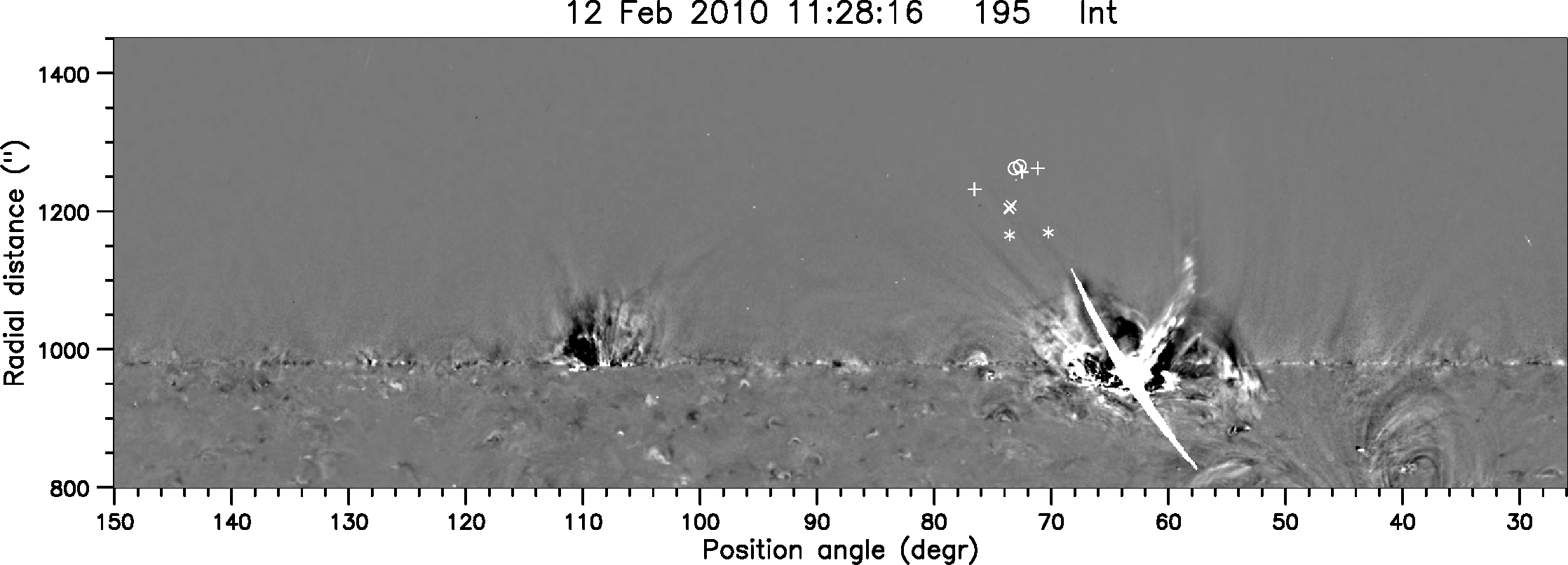}
\includegraphics[width=\hsize]{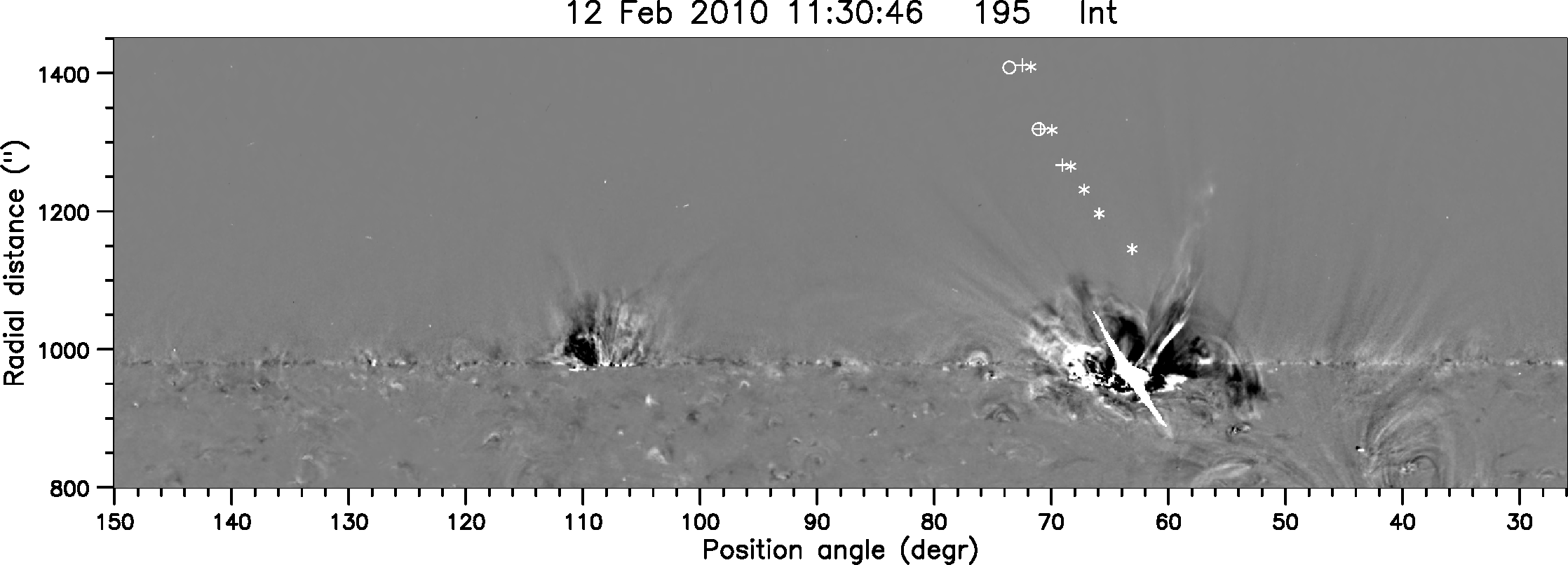}
\caption{Projected positions of the type II radio sources, on top of polar projections of running difference images from 195\,\AA\ STEREO-A. Top: type II during the first phase of the event (11:27:00-11:28:05 UT); asterisks show the average positions from 360.8\,MHz, ``x'' from 298.7\,MHz, crosses from 298.7\,MHz and open circles from 150.9\,MHz. Bottom: type II during the second phase of the event (11:29:21-11:30:57 UT); asterisks are for branch A, crosses for branch B1 and open circles for branch B2. }
\label{195+typeII}
\end{figure}

\begin{figure*}
\centering
\includegraphics[width=.8\hsize]{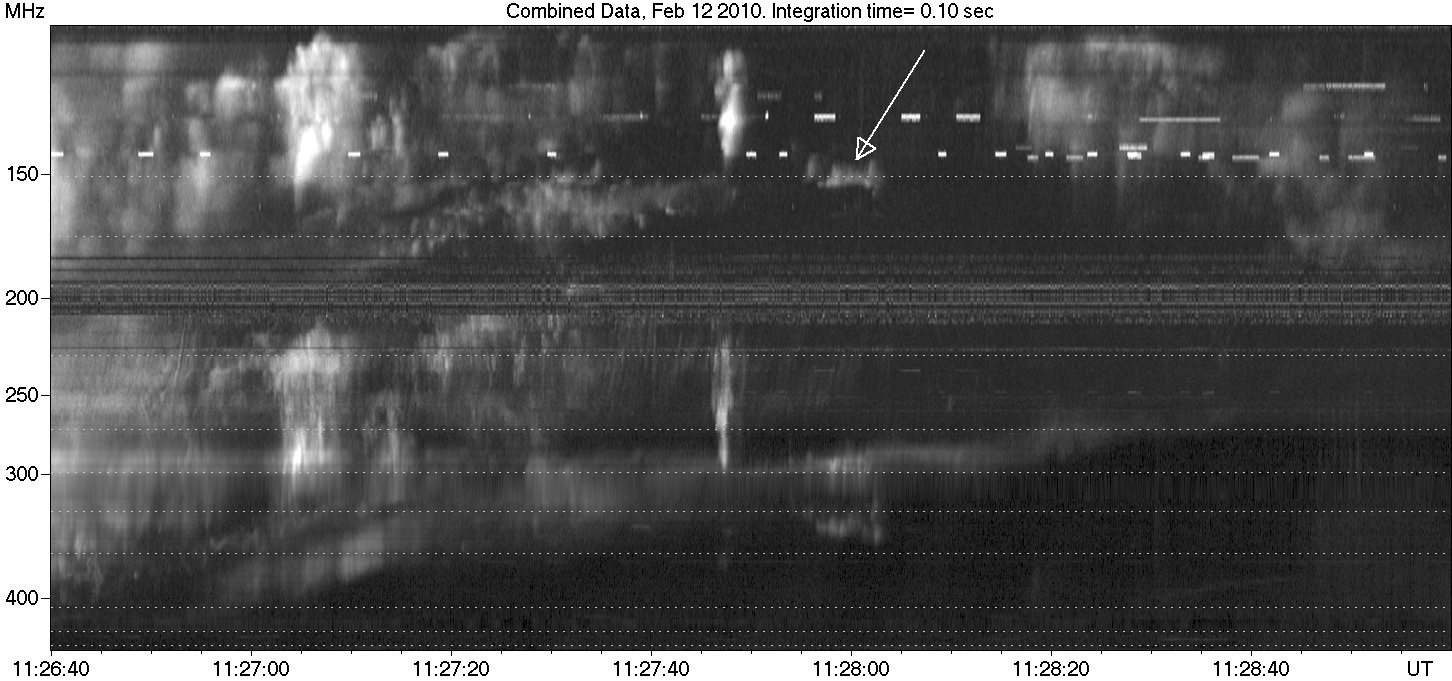}
\caption{Fundamental-harmonic emission during the first phase of the event (composite ASG/SAO dynamic spectrum). The frequency range is from 106 to 450\,MHz. Dotted horizontal lines mark the NRH frequencies. Channels around 200 and 310\,MHz have been partly corrected for radio interference. The arrow points to a spectral feature discussed in the text.}
\label{fig:FH}
\end{figure*}

It is customary in solar metric radio astronomy to associate the emission frequency with the source height and thus obtain 3-dimensional positions. There are some assumptions inherent to this, first of all that the emission is at the plasma frequency or its harmonic, which is true in the case of type II emissions; moreover, an atmospheric model is required and, more seriously it is assumed that the corona is not disturbed, which is probably hardly the case in our situation.EUV waves are associated with small density and temperature enhancements, \emph{e.g.}  see the reviews of \citetads{2009ApJ...700L.182P}; \citetads{2015LRSP...12....3W}; \citetads{2017SoPh..292....7L}, and references therein.  Our EUV data do not contain high-cadence images at multiple channels to perform differential emission measure analysis and unlock the thermal distribution of the plasma perturbed by the EUV wave. An idea about the degree of disturbance of the corona can be obtained from the properties of the EUV wave and the band-splitting. As in almost all cases, our EUV wave is associated with weak (less than 20-30\%) intensity enhancement and hence, possibly, even smaller density enhancements. On the other hand the compression ratio deduced from band-splitting (see Sect.~\ref{MagField} below) is 1.5. 

Having said the above, we obtained three-dimensional positions by assuming an isothermal corona at $2\times10^6$\,K and a base electron density twice that of the Newkirk model. This gave radial distances from 1.21$\cdot$R$_\odot$ (408.0\,MHz, harmonic) to 1.51$\cdot$R$_\odot$ (228.0\,MHz, harmonic) and 1.35$\cdot$R$_\odot$ (150.9\,MHz, fundamental), corresponding heights of 204\arcsec, 495\arcsec\ and 340\arcsec\ respectively.

Our next step was to compute the heliographic coordinates of the sources and from those their projected positions on the nearest 195\,\AA\ STEREO-A images. The results are shown in Fig.~\ref{195+typeII}. In spite of the scatter of the source positions during the first type II (left panel of Fig.~\ref{typeII+jet}), here their projections are close to one another, near the southern top of the 195\,\AA\ EUV wave, but far from the surge. For the second type II the positions of all three branches line up, as if B1 and B2 which started 20\,s after A, followed the same path. These are far from the surge, too; we note, however, that the extrapolation of branch A positions passes close to the disrupted loop system mentioned in Sect.~\ref{Overview}, located north of the surge at position angle of 56\degr.

\begin{figure}[h]
\centering
\includegraphics[width=\hsize]{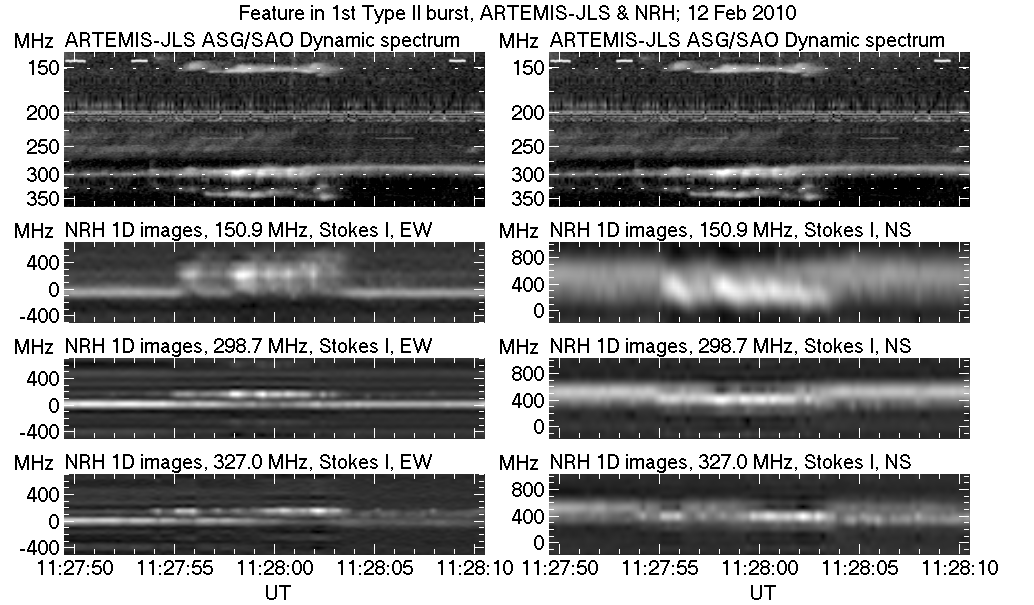}
\caption{Expanded view of the spectral feature marked by the arrow in Fig.~\ref{fig:FH} (top panel, same for both columns). The lower rows show 1D NRH images at three frequencies in the EW (left) and NS (right) directions.}
\label{fig:feature}
\end{figure}

\subsection{Detailed study of the type II bursts}
In addition to the overall evolution of the event described in the previous sections, some particular features are worth mentioning and analyzing; these will be discussed in the present section, together with an in-depth study of the burst properties.

\subsubsection{An impressive fundamental-harmonic structure}\label{fh}
The first interesting feature is a fundamental-harmonic emission (F/H) structure observed during the first phase of the event, already mentioned in Sect. \ref{Overview}, and shown in more detail in Fig.~\ref{fig:FH} where we have combined ASG and SAO data. We note that, in addition to the type II emission, other emissions showed F/H structure as well. 

\begin{figure*}
\centering
\includegraphics[width=.8\textwidth]{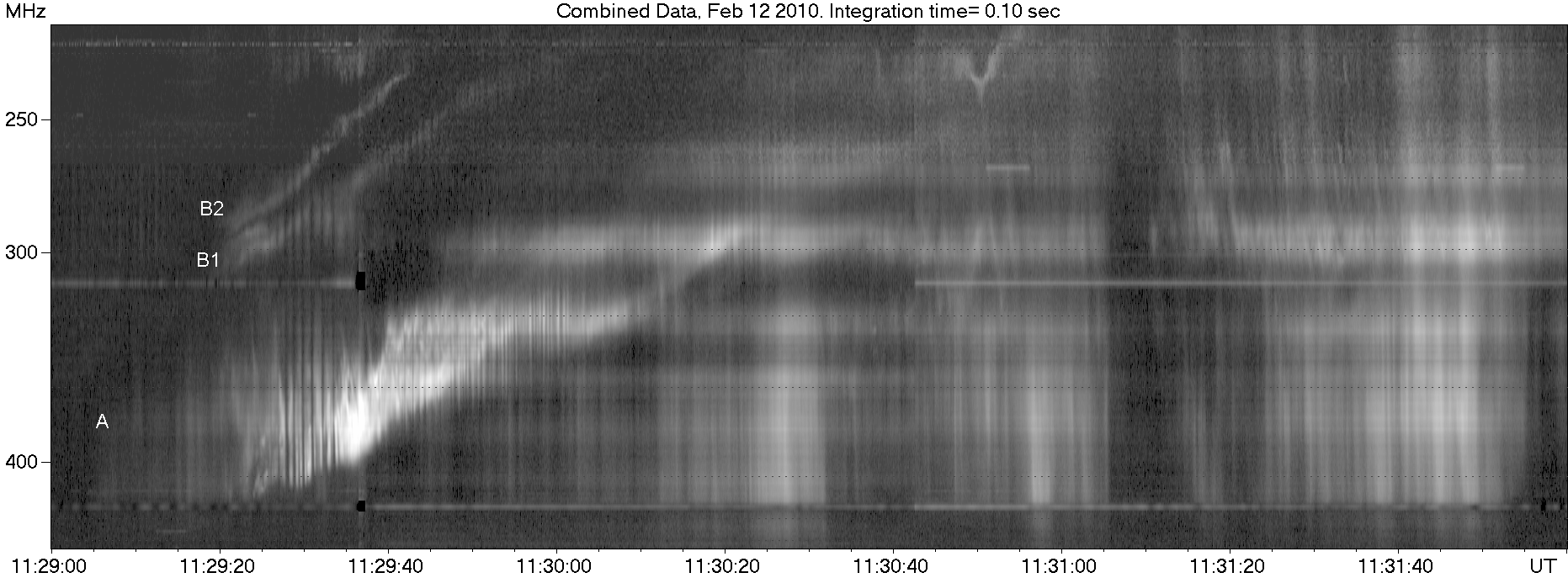}
\caption{Composite ASG/SAO dynamic spectrum of the second type II burst.}
\label{IIb}
\end{figure*}

The type II itself showed band split, to be discussed further in Sect.~\ref{MagField}. It had prominent fine structures at the fundamental, reminiscent of spike-like emissions reported  by \citetads{2019A&A...624A..76A}. The harmonic emission appeared more diffuse, the high frequency split-band in particular, which showed some pulsating emission. Emission at the fundamental was not visible after 11:28:05 UT, while the harmonic emission extended up to 11:28:40 at least.

We employed 2-dimensional cross-correlation to determine the time and frequency shift between the F/H emissions. For the full event we found a frequency ratio of 1.95 and a 0.6\,s delay of the fundamental with respect to the harmonic. However, these values reflect mostly emissions outside the Type II, which are the brightest. Isolating the type II in the dynamic spectrum as much as possible, we obtained a similar frequency ratio and a smaller time delay: 0.23\,s for the entire type II and 0.5\,s for the prominent feature at 11:28 UT (arrow in Fig.~\ref{fig:FH}). 

The measured F/H frequency ratio is consistent with previous works (\citeads{1954AuJPh...7..439W}; \citeads{1963JGR....68.1347M}; \citeads{1959AuJPh..12..327R}); it is expected to be slightly less than 2. Since radiation at the plasma frequency will not propagate, what is observed at the fundamental should originate in the short wavelength wing of the emission profile. We note, however, that in the presence of magnetic field the cutoff frequency of the extraordinary goes above the plasma frequency while that of the ordinary   so that ({\emph{c.f.} eq. 2.122 in \citeads{1996ASSL..204.....Z}):
\be
\upsilon=1\mp\sqrt{u}
\ee
with $\upsilon=(f_{pe}/f)^2$ and $(u=f_{ce}/f)^2$ are parameters expressing the electron density and the magnetic field through the plasma frequency and the electron gyrofrequency respectively. For example, with a magnetic field of 18\,G (see Sect.~\ref{MagField} below) and an observing frequency of 150\,MHz, we have $\sqrt{u}=0.34$ and the cutoff frequency for ordinary radiation is about $1.15f_p$; this will facilitate the escape of ordinary mode radiation at the fundamental.

The time delay is attributed to the difference in the group velocity of the electromagnetic waves near the plasma frequency and its harmonic and to the different paths of the radiation due to refraction and scattering. \citetads{1959AuJPh..12..327R} was the first to measure time delays of the order of 1\,s, from data with a time resolution of 0.5\,s, whereas \citetads{1963JGR....68.1347M} reported delays in the range of 2-150\,s. Here, with an effective time resolution of 0.1\,s and the use of cross-correlation, the delay can be more accurately measured. 

The feature observed in the DS around 11:28 UT, marked by the arrow in Fig.~\ref{fig:FH}, appears to be part of the type II. However, 1D NRH images (Fig.~\ref{fig:feature}) show clearly that it was not, as it was located at a different position, west of the type II source. It lasted for about 8\,s and showed strong intensity fluctuations. Its polarization was of the same sense as that of the type II, but weaker (of the order of 20\%). As a whole it showed practically no frequency drift and had a stronger band split than the main event, visible only in the harmonic. Its role in the event is not clear, yet we considered that it was worth mentioning; its nature would not have been revealed without the imaging information.

\subsubsection{Pulsations during the second type II burst}\label{puls}
Fig.~\ref{IIb} shows the DS of the second type II burst in more detail. As mentioned in Sect.~\ref{Overview}, it consisted of three branches, a wide branch (A in the figures) and two narrow ones (B1 and B2 in the figures). Apart from that, the DS shows superimposed pulsations, an expanded view of which, with five times better time resolution, is shown in the top panel of Fig~\ref{PulsFig}. We note that the pulsations are seen in both the DS and the NRH 1D images; most interesting, they are not limited inside the branches but extend over the continuum emission in between, as clearly shown in the bottom panel of the figure.
Thus the bandwidth of the pulsations was about 200\,MHz.

\begin{figure}[h]
\centering
\includegraphics[width=\hsize]{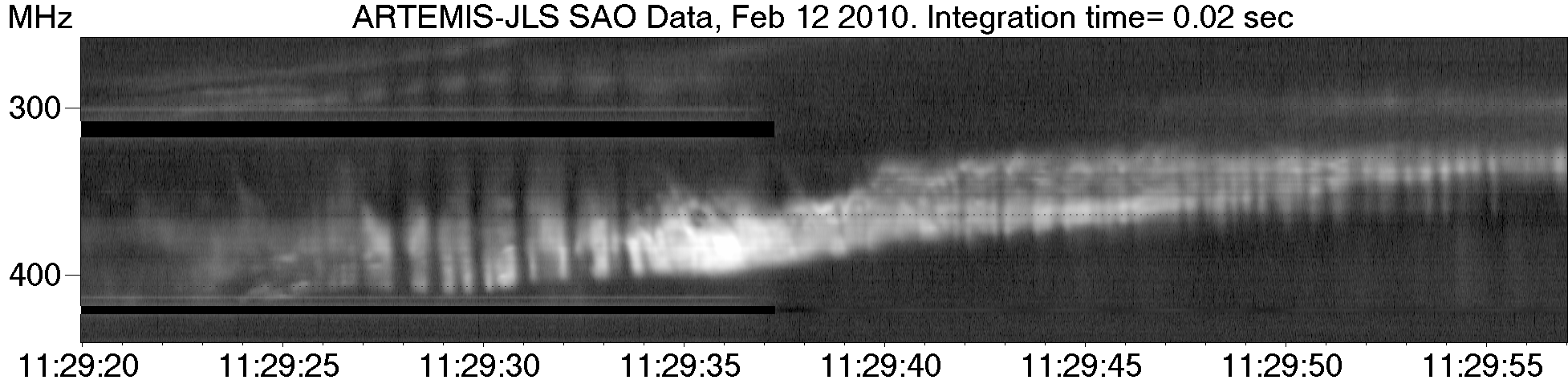}\vspace{.2cm}
\includegraphics[width=\hsize]{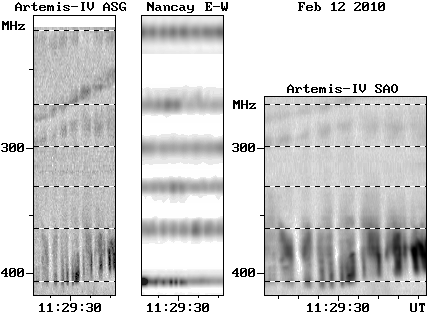}
\caption{Top: Expanded view of the pulsations near the start of the second phase of the event. Bottom: DS of the first part of the pulsations recorded with the ASG (left) and the SAO (right), together with 1-dimensional EW images from the NRH placed at the corresponding DS frequencies (middle). The frequency range is 220-430\,MHz; time tickmarks  are every 2\,s in the left and middle panels, every 1\,s in the right panel. }
\label{PulsFig}
\end{figure}
 
We note that the period of pulsations decreased with time; a power spectrum analysis gave an initial period of 1.13\,s, dropping progressively to 0.42\,s. These values are within the range reported in the literature   (\emph{e.g.} \citeads{2007LNP...725..251N}; \citeads{2015SoPh..290..219B}). It is also noteworthy that in the initial pulsations there is embedded temporal fine structure (right bottom panel of Fig.~\ref{PulsFig}) and that the duration of their high intensity phase is longer than that of the low intensity phase.  

A detailed examination of the 1D NRH images revealed that, not only the pulsating source existed before the start of the second type II, but there was also a second source of opposite circular polarization (Fig.~\ref{Puls_before}), which started during the first phase of the event, around 11:28:46 UT. The two sources were visible in all NRH frequencies between 228.0\,MHz and 408.0\,MHz, but hardly visible in the DS due to their low intensity and phase differences in the intensity fluctuations which reduced the overall signal. The source with negative polarization disappeared near the start of the type II, and it is tempting to speculated that the shock might be the result of the interaction of the two pulsating regions.

\begin{figure}
\centering
\includegraphics[width=\hsize]{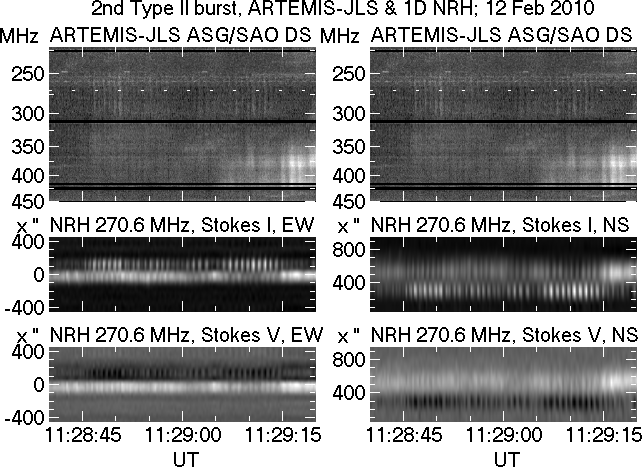}
\caption{Pulsations before the start of the second type II; Dynamic spectrum (top) and 1D NRH images at 270.6\,MHz in Stokes I and V, EW and NS.}
\label{Puls_before}
\end{figure}
 
Pulsations are typical of type IV bursts. In our case, there is certainly a type IV continuum background; although this is not clearly
visible in the DS due to the rich superposed fine structure, it is visible between the branches of the second burst as mentioned above. It appears that whatever caused the background pulsations induced pulsations in the emitting shock fronts as well; an alternative interpretation is that the continuum emission region is located in front of the front and is optically thin, so that its pulsations modulated the intensity of the type II emission.  

We used cross-correlation of the average intensity of selected DS spectra channels with all other channels in an effort to measure frequency drifts. We found a reverse relative drift of about 2\,s$^{-1}$; such a high drift cannot be interpreted in terms of motions, so that its origin needs further investigation.

\subsubsection{Drift rates and velocity estimates}\label{DriftRates&Velocity}
The drift rate in the early part of the first type II was $-7.7\times10^{-3}$\,s$^{-1}$; around 11:27:20 UT the drift decreased and the first type II appeared stationary until 11:28:05 UT (see Fig.~\ref{fig:FH}), at which time a small jump to lower frequencies was observed, followed by another jump around 11:28:20 UT. Thus the value of the average drift over the $\sim110$\,s duration of the burst dropped to $-3.3\times10^{-3}$\,s$^{-1}$ (Table~\ref{tab:drift}).

Assuming plasma emission, the frequency drift is related to the velocity of the exciter, $\upsilon$ and the scale length of the electron density, $L_{Ne}$ along the path of the exciter. $\ell$:
\bea
\frac{1}{f}\frac{df}{dt}&=&\frac{1}{2}\,\upsilon\,L_{Ne}^{-1}, \hspace{.6cm}\mbox{and}\\ \label{eq:drift} 
L_{Ne}^{-1}&=&\frac{1}{N_e}\frac{dN_e}{d\ell}
\eea

\begin{table}[h]
\begin{center}
\caption{Type II drift rates and nominal velocities}
\begin{tabular}{lcc}
\hline
Burst                             & Drift & Velocity\\
                                    & $10^{-3}$\,s$^{-1}$ & km\,s$^{-1}$\\
\hline
Start of first type II         & $-7.7$ & 1540 \\
Average of first type II    & $-3.3$ & 660  \\
Branch A, low freq edge  & $-17$  & 3400 \\
Branch A, high freq edge & $-9.2$ & 1840 \\
Branch B1                      & $-8.8$ & 1760 \\
Branch B2                      & $-12$  & 2400 \\
\hline
\label{tab:drift}
\end{tabular}
\end{center}
\end{table}

Classically, the drift is used to measure $\upsilon$, assuming a constant value for $L_{Ne}$. However, in the present case of variable drift and the complex coronal environment, the observed decrease of the drift could be attributed to a decrease of the shock speed and/or to an increase of the density scale length. Moreover, the value of $L_{Ne}$ is usually assumed to be equal to the scale height predicted by a particular spherically symmetric coronal model, so that the computed value of $\upsilon$ represents the radial component of the exciter velocity. With coronal models being practically isothermal at a temperature $T$, the scale height at a radial distance $r$, $H(r)$, is given by:
\be
H(r)=H_\odot \frac{r^2}{R_\odot^2}=\frac{kT}{\mu g_\odot m_{\rm{H}}} \frac{r^2}{R_\odot^2}
\ee
where $H_\odot$ is the scale height at $r=R_\odot$, $\mu$ is the average molecular weight and $g_\odot$ the gravity at $r=R_\odot$. Thus the decrease of the drift rate might indicate that the shock encountered a high-temperature region with increased density scale length.
 
The above discussion shows that there are many uncertainties in computing the shock speed from the frequency drift. Nevertheless, we added the value of this nominal velocity, computed for a typical value of $H=10^5$\,km, in Table~\ref{tab:drift}.

The bandwidth of branch A in the second burst increased with time, so that its low frequency edge had a higher drift rate 
than its high frequency edge (Table~\ref{tab:drift}). Branches B1 and B2 were short ($\sim30$\,s) and narrowband, 7-14\,MHz wide (relative bandwidth of 2.7-5.3\%), with drifts between those of the two edges of branch A. No significant time variations of the drift rate were detected in this burst. We note that the drift rates of the components of the second type II were up to a factor of 2 higher than the drift at the start of the first type II and that all nominal velocities are considerably higher than the velocities derived from WL and EUV images (Table~\ref{tab:vel}).
 
In the case of  branch A of the second burst we had position measurements at a sufficient number of frequencies to obtain a 3D trajectory by associating the emission frequency to the source height (Sect.~\ref{positions} and Fig.~\ref{typeII+jet}). Using this information we computed a radial velocity of $2450\pm260$\,km\,s$^{-1}$. This value is very close to the average velocity of the low and high frequency edges of the branch, derived from the frequency drift. From the same trajectory we computed a horizontal velocity of $2050\pm230$\,km\,s$^{-1}$, directed southwards, and a total velocity of $3200\pm230$\,km\,s$^{-1}$.

\subsubsection{Magnetic field from band splitting}\label{MagField}
Band splitting is widely accepted to be due to the density jump at the shock front, {\it i.e.} between the uncompressed plasma in front of the shock and the compressed plasma behind it. From the frequency ratio of the two bands the Alfv\'en Mach number, $M_A$ can be computed and from that, together with the density scale and the frequency drift, the intensity of the magnetic field can be estimated (\citeads{1974IAUS...57..389S}; \citeads{2002A&A...396..673V}, see also the review by \citeads{2021FrASS...7...77A}). 

Combining (\ref{eq:drift}) with the definition of the Alfv\'en velocity we get for the magnetic field, $B$:
\be
B= 4\sqrt{\pi N_e m_p} \frac{L}{M_A} \frac{1}{f}\frac{df}{dt}
\ee
where $M_A$ is the Alfv\'en Mach number which, according to Vr\v{s}nak \etal\ 2002, can be computed from:
\be
M_A= \sqrt{\frac{X(X+5)}{2(4-X)}}
\ee
with $X$ being the square of the band-split frequency ratio.

During the initial phase of the first burst we measured a band-split frequency ratio of $1.23\pm0.06$, which corresponds to a density ratio of 1.5; applying the above method, we obtained  $M_A=1.18\pm0.05$ and a magnetic field strength of $18.4\pm0.7$\,G for a $2\times10^6$\,K isothermal coronal model with base density $2\times$ the Newkirk model value. Here the errors refer to the rms of values measured in the DS at different frequencies/times. We note that the empirical relation of \citetads{1978SoPh...57..279D} predicts $B=2.8$\,G at $r=1.32$\,R$_\odot$, which is the average radial distance to which our measurements refer, according to the model used. Our value is also higher than values reported in the literature, see \citetads{2021FrASS...7...77A}. This difference could be attributed to the peculiarities of the event.

\subsubsection{Polarization of the emission}
We measured the circular polarization (Stokes parameter $V$) of the type II burst from the NRH images. For the first type II we found positive (right hand) polarization for both the harmonic and the fundamental emission. Taking into account the discussion in Sect.~\ref{fh}, the fundamental is polarized in the sense of the ordinary mode; the fact that the harmonic has the same sign of $V$ implies that it is also polarized in the ordinary mode sense. The degree of polarization was measured to be about 35\% for the fundamental and 65\% for the harmonic. Normally the fundamental should be 100\% polarized, but could be depolarized as a result of scattering and/or propagation effects (see Sect. 6.1 in \citeads{2021FrASS...7...77A}). The second type II burst, presumably emitting in the harmonic, was also polarized in the positive sense, the polarization degree ranging from 50\% to 90\%. We add that, with the exception of the pulsating source mentioned in Sect.~\ref{puls}, the entire event was polarized in the right hand sense.

In the literature there are diverse reports about type II polarization, ranging from no polarization to strong polarization. \citetads{1958AuJPh..11..201K} and \citetads{1959AuJPh..12..327R} reported weak or no polarization, whereas \citetads{1987HvaOB..11..111Z} reported cases of both weak and strong polarization, noting that fine structures (herringbone) were more polarized than the burst continuum.  \citetads{1987SoPh..111..365C} reported 15\% polarization for herringbones, with the harmonic exhibiting weaker polarization than the fundamental.  \citetads{2014ApJ...795...14H} also reported low polarization (5-10\%) in the harmonic for a single event, while \citetads{2014ApJ...793L..39D} reported 30-60\% for the fundamental in another single event.  \citetads{2003ApJ...592.1234T} gave 50-90\% polarization for a fragmented type II burst at high frequencies, while strong polarization at the fundamental and weak for the harmonic was reported by \citetads{2001JGR...10625313T} for interplanetary type IIs.

\section{Discussion: origin of type II emission}\label{Discuss}

The two type II bursts were associated with an M9 flare and occurred in close time succession around the peaks of the two phases of the flare. Due to their timing with respect to the flare evolution, it is tempting to attribute their origin to pressure pulses generated by the flare. Our data did not allow us to perform a high-cadence emission measure analysis that could have directly revealed the possible existence of pressure pulses ignited by the flare and their relevance to the generation of the type II bursts. \citetads{2015LRSP...12....3W}  pointed out that flares igniting coronal waves are usually located at the  periphery of their active regions and the waves are launched into directions away from the core of the active regions; this pattern facilitates the development and steepening of a pressure pulse since regions of low magnetic field strength and low Alfv\'en speeds are accessed. In our case the flare indeed occurred away from the core (defined as the location of the main polarity inversion line) of the active region (\emph{e.g.} see panels 30-35 of Fig. \ref{figure2}) but the positions of the first type II sources (see Fig. \ref{195+typeII}) are at odds with Warmuth's (\citeyearads{2015LRSP...12....3W}) suggestion.

The positions of both type II bursts are broadly consistent with the EUV wave whose generation and development are presented in Fig. \ref{figure3}.Other case-studies showing spatio-temporal association between EUV waves and metric type II  bursts include \citetads{2014SoPh..289.2123K}; \citetads{2019A&A...624L...2M}; \citetads{2020A&A...639A..56J}; \citetads{2020A&A...644A..90K}. There has been extensive literature about the drivers of EUV waves (see the reviews by \citeads{2012SoPh..281..187P}; \citeads{2015LRSP...12....3W}; \citeads{2017SoPh..292....7L}, and references therein). As in most cases, the lateral propagation of our EUV wave makes it difficult to attribute its generation to the flare \citepads[\emph{e.g.}][]{2009ApJ...700L.182P}. If we exclude the flare, we need an alternative low coronal driver in the form of mass motion. The only reliable candidate is the surge; from the timeline of the event it appears that the surge generated the disturbance that yielded the EUV wave. An example of an EUV wave linked to a surge has been presented by \citetads{2013ApJ...764...70Z}. Generally speaking, narrow eruptions (surges, jets etc) could drive EUV waves (\emph{e.g.}\citeads{2015LRSP...12....3W} and references therein).

This is somehow different from the scenario of \citetads{2012SoPh..281..187P} which identifies the lateral over-expansion of the CME cavity in the low corona as the EUV wave driver. However, our EUV data from both STEREO spacecraft do not show concrete evidence of a cavity. The same applies to the WL observations of the associated CME with  COR1. It is possible that the line-of-sight was not favorable in order to reveal a complete CME cavity by, for example, not being aligned with its axis and/or not having enough mass along the line of sight to produce appreciable signal in EUV or WL (see also \citeads{2013SoPh..284..179V}, for a detailed discussion on how projection effects affect the appearance of CMEs). We therefore conclude that the scenario is not directly applicable to our event.
 
Evidence for the propagation of the disturbance was also provided by the streamer disruption while its evolution  was detected as a white-light CME in the coronagraph data.

The loop system located at a position angle of 56\degr\ north of the surge (see Sect. \ref{positions}) was most probably disrupted by the wave but was not the driver of the wave. The latter conclusion is also supported by the following arguments: (1) At 11:28 UT (that is, around the start of the second phase of the event) the STEREO A images indicate that the EUV wave has propagated well beyond these loops to the north. (2) The disrupted loops expand along a direction which is markedly different from the direction defined by the positions of the second shock (see Fig. 8, bottom). (3) The amplitude of the loop oscillations appears too small to give rise to a major disturbance such as the one associated with the second shock.

Our results directly link the same EUV wave with the origin of both type II bursts. This conclusion is different from the traditional interpretation of the origin of successive type II bursts in the absence of successive CMEs which invokes a CME driver for one burst and a flare-related origin for the other (\emph{e.g.} \citeads{2005SoPh..232...87S}; \citeads{2006A&A...451..683S}). On the other hand, the origin of multi-lane type II bursts have occasionally been interepreted as coming from distinct locations of a single CME-driven shock 
(\emph{e.g.} \citeads{2015SoPh..290.1195F}; \citeads{2017SoPh..292..194L}). 

In the light of the association of the EUV wave with the type II bursts a few remarks are in order: 

(1) The association of two separate type II bursts with the same EUV wave indicate that each type II is related with different parts of the wave; supporting evidence for this argument is primarily provided by the positional offsets between the sources of the two type II bursts and by the different rate of expansion across the wave. The three-dimensional structure of the Alfv\'en speed above the active region as well as the magnetic field configuration that would favor the acceleration of electrons at distinct locations of the shock front might have also contributed to the appearance of the type II sources at the observed positions. Unfortunately our data were not suitable for a detailed study of these latter effects.

(2) The speed of both type II bursts computed from either the drift rates measured in the dynamic spectra or from the projected positions of the  radio sources are systematically higher than the driver (that is, the surge which generated the EUV wave) speed (see Section \ref{DriftRates&Velocity}). Combining the speed at the start of the first type II burst (see Table \ref{tab:drift}) and the value of the Alfv\'en Mach number found in Section \ref{MagField} we find an Alfv\'en speed of about 1300 km s$^{-1}$. Therefore the driver propagates at sub-Alfv\'enic speeds. This situation is indicative of a piston-driven shock \citepads[\emph{e.g.} see][]{2008SoPh..253..215V}, during the first stages of the event. However, the movies suggest that after some point in time the EUV wave attained large  lateral distances from the surge,and therefore, the wave (shock) was then possibly freely-propagating.

\begin{figure}
\centering
\includegraphics[width=.8\hsize]{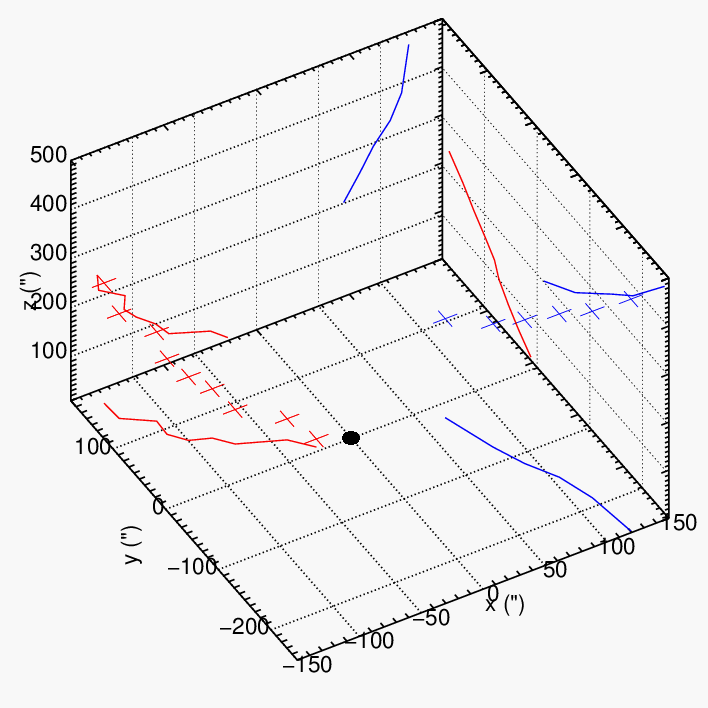}
\caption{3D rendering of the surge and NRH source positions for branch A of the second type II burst (blue and red crosses respectively), together with their projection on the basic planes, xy, xz and yz (blue and red lines). The black dot marks the position of the flare.}
\label{3D}
\end{figure}

\section{Summary and conclusions}\label{Concl}
In this work we used all available data, from km radio waves to hard X-rays, in order to study the event SOL2010-02-012T11:25:00 of February 12, 2010 in conjunction with dynamic spectra from ARTEMIS-IV/JLS and images from the NRH. Our emphasis was on the origin of the two metric type II bursts.

The low cadence of the non-radio data made difficult the association of the fast evolving metric emission with the phases of the event in other parts of the electromagnetic spectrum. Still, the NRH observations near the central meridian reduced the influence of refraction and scattering on the source positions, whereas observations from STEREO A and B provided three-dimensional information about the event. These gave to our study some important advantages compared to other similar works.

Based on computations for homogeneous models, we did not apply any refraction/scattering corrections to the position of the radio sources because the resulting corrections were too small to affect our conclusions. However, this issue warrants further investigation as has been done for the decametric wavelength range  \citepads[\emph{e.g.}][]{2021ApJ...909..195Z}; it is also important to investigate what happens in the (normal) inhomogeneous case. In addition to refraction/scattering, we ignored ionospheric effects which, according to our measurements, were considerably smaller than the size of the radio sources.

The complex event originated in a very compact flare, and was accompanied by a surge, a CME, a streamer disruption and an EUV wave. Our analysis, discussed in detail in Sect.~\ref{Discuss}, showed that the two associated type II bursts were not related either with the surge (see Fig.~\ref{3D}) or with the streamer disruption, but with the EUV wave which was probably driven by the surge. We could not identify any non-thermal radio signatures of the surge or the streamer disruption; it is possible, however, that such signatures might be buried deep into the rich structure of the dynamic spectrum. Emission from the surge was detected at a later phase of the event, and this will be presented in a subsequent publication.

It is also important to discuss the possible effect of the adopted coronal model to the derived NRH source positions, through the model dependence of the emission height. To this end we computed heights for two more isothermal models, one (model 2) with a coronal temperature of $2\times10^6$\,K and a base density equal to that of the Newkirk model and another (model 3) with $T=1.4\times10^6$\,K and a base electron density twice that of the Newkirk model. Although both models gave smaller source heights, they did not bring the radio sources any closer to the surge. As for the speed derived from the frequency drift for the second type II, model 2 gave a value close to that of the adopted model whereas model 3 gave a 40\% lower value, still more than a factor of 2 above the WL/EUV measurements.

In addition to the origin of the type II emission, our analysis of the radio observations gave a number of interesting results which are worth of further investigation:

Using cross-correlation we measured the frequency ratio and the time delay between emission at the plasma frequency and its harmonic during the first type II burst; we found values of 1.95 and 0.23-0.6\,s respectively. This is a powerful technique that can be used systematically to derive precise values of these parameters and thus provide more accurate information on the formation of the fundamental as well as on wave propagation effects. 

The first burst started with a relatively high frequency drift; subsequently the drift diminished to increase again later on. \citetads{2020ApJ...893..115C} reported a case of transition of a type II from stationary to drifting, but we are not aware of any report of drifting-stationary-drifting transition. Such a transition might be due to the shock encountering regions of different density scale lengths.

For the same burst we inferred a magnetic field of 18\,G from band splitting, a rather high value compared to those reported in the literature. We note that the derived magnetic field is model-dependent, dropping to 14\,G for model 2 (see above) and to 11\,G for model 3. 

We observed pulsations superposed on the second type II burst. The pulsations were broad-band  with a bandwidth of about 200\,MHz and they started before the type II. They had a measurable (reverse) drift of about 2\,s$^{-1}$. Moreover, thanks to the NRH images, we identified a second pulsating source of opposite circular polarization which disappeared near the start of the burst. This is a potentially important finding, since we are not aware of any other report of multiple pulsation sources, but its full investigation is outside the scope of this work. 

Finally, we would like to stress that this work confirms the importance of imaging spectroscopy for understanding the physics of solar radio emission.

\begin{acknowledgements}
The authors are grateful to the other members of the ARTEMIS group, in particular to C. Caroubalos, P. Preka-Papadema and X. Moussas (University of Athens), as well as to A. Kontogeorgos and P. Tsitsipis (University of Thessaly). We also wish to thank L. Klein and the entire group of the Nan\c cay Radioheliograph for the original NRH visibility data, G. Chernov for interesting discussions and E. Kontar for the scattering computations. Data from STEREO (SECCHI and coronagraphs), SXI, Hinode, the Catania observatory, RHESSI, and GOES were obtained from the respective data bases; we are grateful to all those who contributed to the operation of these instruments and made the data available to the community. Finally we wish to thank the Onassis Foundation for financial support (Grant 15153) for the continued operation of the ARTEMIS-JLS radio spectrograph. 
\end{acknowledgements}

\bibliographystyle{aa}
\bibliography{12Feb2010}

\end{document}